\pdfoutput=1
\documentclass[11pt,twoside,a4paper,cmspaper,final,collab]{cms-tdr}
\def\svnVersion{f20fcdf}\def\svnDate{2024/04/27}\def\cmsCernNoTag{CERN-EP-2023-284}\def\cmsCernDate{\today}\def\cmsMessage{Published in the European Physical Journal C as \href{http://dx.doi.org/10.1140/epjc/s10052-024-12727-4}{\doi{10.1140/epjc/s10052-024-12727-4}.}}
\begin{document}\cmsNoteHeader{HIG-22-007}

\newcommand{\PNfix}{\hspace{-.04em}}
\newcommand{\tautau}{{\ensuremath{\PGt\PNfix\PGt}}\xspace}
\newcommand{\emu}{{\ensuremath{\Pe\PNfix\PGm}}\xspace}
\newcommand{\etau}{{\ensuremath{\Pe\PNfix\tauh}}\xspace}
\newcommand{\mutau}{{\ensuremath{\PGm\PNfix\tauh}}\xspace}
\newcommand{\mtt}{{\ensuremath{m_{\tautau}}}\xspace}
\newcommand{\wjets}{{\ensuremath{\PW\PNfix}+jets}}
\newcommand{\il}{138\fbinv\xspace}
\newcommand{\mfour}{\ensuremath{m_{\PGm\PGm\PQb\PQb}}\xspace}
\newcommand{\Hb}{\PH\xspace}
\DeclareRobustCommand{\Pai}{{\HepParticle{a}{1}{}}\Xspace}
\newcommand{\ma}{\ensuremath{m_{\Pai}}\xspace}
\newcommand{\ab}{\Pai\xspace}
\newcommand{\mmumu}{\ensuremath{m_{\PGm\PGm}}\xspace}
\renewcommand{\mtt}{\ensuremath{m_{\PGt\PGt}}\xspace}
\newcommand{\mbb}{\ensuremath{m_{\PQb\PQb}}\xspace}
\newcommand{\resbb}{\ensuremath{\sigma_{\PQb\PQb}}\xspace}
\newcommand{\resh}{\ensuremath{\sigma_{\Hb}}\xspace}
\newcommand{\mmbb}{\ensuremath{\PGm\PGm\PQb\PQb}\xspace}
\newcommand{\ttbb}{\ensuremath{\PGt\PGt{\PQb}{\PQb}}\xspace}
\newcommand{\chit}{\ensuremath{\chi_\text{tot}}\xspace}
\newcommand{\chih}{\ensuremath{\chi_{\Hb}}\xspace}
\newcommand{\chihl}{\ensuremath{\frac{(\mfour - 125\GeV)}{\resh}}\xspace}
\newcommand{\chibl}{\ensuremath{\frac{(\mbb - \mmumu)}{\resbb}}\xspace}
\newcommand{\chib}{\ensuremath{\chi_{\PQb\PQb}}\xspace}
\newcommand{\chihd}{\ensuremath{\chi_{\Hb,\mathrm{d}}}\xspace}
\newcommand{\chibd}{\ensuremath{\chi_{\PQb\PQb,\mathrm{d}}}\xspace}
\newcommand{\chid}{\ensuremath{\chi_\mathrm{d}}\xspace}
\newcommand{\toppair}{{\PQt{}\PQt}\xspace}
\newcommand{\mellell}{\ensuremath{m_{\ell\ell}}\xspace}
\newcommand{\dzeta}{\ensuremath{D_{\zeta}}\xspace}
\newcommand{\pzeta}{\ensuremath{p_{\zeta}}\xspace}
\newcommand{\ptem}{\ensuremath{\pt^{\Pe/\PGm}}}
\newcommand{\mjj}{\ensuremath{m_\mathrm{jj}}}
\newcommand{\irel}{\ensuremath{I_\text{rel}}}
\ifthenelse{\boolean{cms@external}}{\providecommand{\cmsLeft}{upper\xspace}}{\providecommand{\cmsLeft}{left\xspace}}
\ifthenelse{\boolean{cms@external}}{\providecommand{\cmsRight}{lower\xspace}}{\providecommand{\cmsRight}{right\xspace}}
\ifthenelse{\boolean{cms@external}}{\providecommand{\cmsURight}{middle\xspace}}{\providecommand{\cmsURight}{upper right\xspace}}
\ifthenelse{\boolean{cms@external}}{\providecommand{\cmsULeft}{upper\xspace}}{\providecommand{\cmsULeft}{upper left\xspace}}

\cmsNoteHeader{HIG-22-007}
\title{Search for exotic decays of the Higgs boson to a pair of pseudoscalars in the \texorpdfstring{$\PGm\PGm\PQb\PQb$}{mu mu b b} and \texorpdfstring{$\PGt\PGt\PQb\PQb$}{tau tau b b} final states}

\author*[cern]{The CMS Collaboration}

\date{\today}

\abstract{
A search for exotic decays of the Higgs boson (\PH) with a mass of 125\GeV to a pair of light pseudoscalars $\ab$ is performed in final states where one pseudoscalar decays to two $\mathrm{b}$ quarks and the other to a pair of muons or $\tau$ leptons. A data sample of proton-proton collisions at $\sqrt{s}=13\TeV$ corresponding to an integrated luminosity of 138\fbinv recorded with the CMS detector is analyzed. No statistically significant excess is observed over the standard model backgrounds. Upper limits are set at 95\% confidence level (\CL) on the Higgs boson branching fraction to $\PGm\PGm\PQb\PQb$ and to $\PGt\PGt\PQb\PQb$, via a pair of $\ab$s. The limits depend on the pseudoscalar mass \ma and are observed to be in the range (0.17--3.3)$\times 10^{-4}$ and (1.7--7.7)$\times 10^{-2}$ in the $\PGm\PGm\PQb\PQb$ and $\PGt\PGt\PQb\PQb$ final states, respectively. In the framework of models with two Higgs doublets and a complex scalar singlet (2HDM+S), the results of the two final states are combined to determine upper limits on the branching fraction $\mathcal{B}(\PH\to\ab\ab\to \ell\ell\PQb\PQb)$ at 95\% \CL, with $\ell$ being a muon or a \PGt lepton. For different types of 2HDM+S, upper bounds on the branching fraction $\mathcal{B}(\PH\to\ab\ab)$ are extracted from the combination of the two channels. In most of the Type II 2HDM+S parameter space, $\mathcal{B}(\PH\to\ab\ab)$ values above 0.23 are excluded at 95\% \CL for \ma values between 15 and 60\GeV.
}

\hypersetup{
pdfauthor={CMS Collaboration},%
pdftitle={Search for exotic decays of the Higgs boson to a pair of pseudoscalars in the \texorpdfstring{$\PGm\PGm\PQb\PQb$}{mu mu b b} and \texorpdfstring{$\PGt\PGt\PQb\PQb$}{tau tau b b} final states},
pdfsubject={CMS},
pdfkeywords={CMS, Higgs, exotic, physics}}

\maketitle 

\section{Introduction}
The discovery of the Higgs boson (\PH) by the ATLAS and CMS experiments at the CERN LHC~\cite{Aad:2012tfa, Chatrchyan:2012xdj, Chatrchyan:2013lba} strengthened the case for the standard model (SM), which states that the electroweak (EW) symmetry is broken by a complex scalar field~\cite{PhysRevLett.13.321, Higgs:1964ia, PhysRevLett.13.508, PhysRevLett.13.585, Higgs:1966ev, Kibble:1967sv}.
However, the SM is not a complete theory as it cannot account for a number of experimental observations.
For example, the origin of neutrino mass and dark matter remains unexplained in the SM.
Several beyond the SM (BSM) theories address these observations while identifying the 125\GeV resonance as part of an extended group of scalar particles.
The Two-Higgs-Doublet Models (2HDMs)~\cite{PhysRevD.8.1226, Branco_2012, Djouadi:2005gj} predict five physical scalar and pseudoscalar particles and allow different couplings of each scalar to SM fermions.
The two real scalar singlet extension~\cite{Robens_2015, Robens_2020} of the SM results in three neutral scalar bosons.
A broad class of 2HDMs extended with an additional complex scalar singlet (2HDM+S) contains seven physical scalar and pseudoscalar particles~\cite{Branco_2012}.
In all these models, one of the scalars is identified as the discovered Higgs boson with a mass of 125\GeV.

Recent measurements of the Higgs boson's couplings at the LHC do not rule out exotic decays of the Higgs boson to BSM particles.
The ATLAS and CMS experiments put, respectively, 12 and 16\% upper bounds on the branching fraction of the Higgs boson to undetected particles at 95\% confidence level (CL) using data collected in 2016-2018 (Run~2)~\cite{naturecombatlas, naturecombcms}.
Given these bounds, it is crucial to examine the data for direct evidence of new particles coupling to the Higgs boson, in particular, to test possible extensions of the SM.

The exotic decay channels may include the Higgs boson decaying to a pair of light pseudoscalar particles that subsequently decay to pairs of SM particles.
This signal can be experimentally discriminated from SM Higgs boson decays.
These decays arise naturally in the phenomenology of 2HDM+S, which is described here in more detail.
The 2HDM+S couplings are such that a fermion can couple to only one of the scalar doublets to avoid flavor changing neutral currents at tree level.
Under this condition, four types of 2HDM+S models are possible~\cite{Branco_2012, PhysRevD.90.075004}.
While the SM-like couplings of the Higgs boson to fermions and gauge bosons can be preserved, the singlet state of the 2HDM+S can also serve as a dark matter candidate that couples to the Higgs boson~\cite{Grzadkowski_2010, Drozd_2014}.
In 2HDM+S scenarios of Type I, the second doublet, $\phi_2$, can couple to any fermion whereas the first doublet, $\phi_1$, cannot couple to fermions.
In Type II models, $\phi_1$ couples to down-type quarks and charged leptons while $\phi_2$ couples to up-type quarks.
This model is close to the next-to-minimal supersymmetric SM (NMSSM), which is a special case of 2HDM+S and provides a solution to the so-called $\mu$-problem~\cite{KIM1984150, muproblem}.
The NMSSM particle spectrum contains two pseudoscalars, \ab and $\mathrm{a}_2$, the lighter \ab can have a mass smaller than the Higgs boson to allow $\Hb\to\ab\ab$ decays.
Another valid extension has quarks coupling to $\phi_2$ and charged leptons coupling to $\phi_1$, referred to as the Type III or ``lepton-specific'' model.
Finally, in the Type IV or ``flipped'' model, $\phi_2$ couples to up-type quarks and charged leptons while $\phi_1$ couples to down-type quarks~\cite{Branco_2012, PhysRevD.90.075004}.

The branching fraction, $\mathcal{B}$, of $\ab\ab\to$ SM particles depends on the type of 2HDM+S model, the mass of the pseudoscalar, \ma, and the ratio of the vacuum expectation values of the two doublets, \tanb.
The decay width of \ab to fermion pairs depends, in addition, on the mass of the decay products.
In Type II 2HDM+S models, $\mathcal{B}(\ab\ab\to\ttbb)$ is slightly above 10\%, while it can reach up to $\sim$50\% in Type III models.
The large predicted branching fraction makes this channel particularly attractive.
The decay of \ab pairs to \mmbb has a much smaller branching fraction.
In Type III models, for $\tanb=2$, $\mathcal{B}(\ab\ab\to\mmbb)$ is predicted to be about 0.2\%.
Despite the small branching fraction, this channel can provide competitive results given the high performance of the muon reconstruction and the excellent dimuon mass resolution in CMS.
The possibility of the Higgs boson decaying into a pair of $\ab$s is studied in this paper for both \ttbb and \mmbb decay modes.
The gluon-gluon fusion production mechanism (${\Pg\Pg}$F) constitutes the dominant Higgs bosons production process, with a cross section of $\sigma_{\rm ggF}^{13 \TeV}=48.58\pm 1.56\unit{pb}$~\cite{deFlorian:2016spz} at next-to-next-to-next-to-leading order ($\mathrm{N^{3}LO}$) accuracy in perturbative quantum chromodynamics (QCD) and next-to-leading order (NLO) in EW corrections.
The contribution of the Higgs boson production through vector boson fusion (VBF) is also taken into account with a cross section of $\sigma_{\PQq\PQq\Hb}^{13 \TeV}=3.72\pm 0.08\unit{pb}$~\cite{deFlorian:2016spz}, which includes NLO QCD and EW corrections.

Similar searches have been performed at the LHC.
The latest analysis by the ATLAS Collaboration~\cite{ATLAS:2021ypo} has placed a strong bound of $\mathcal{B}(\Hb\to\ab\ab\to \mmbb)< (0.2$--$4)\times 10^{-4}$ in the range $16<\ma<62\GeV$, using the LHC Run~2 data at $\sqrt{s}=13\TeV$, extending its prior analysis with a partial data sample~\cite{Aaboud:2018esj}.
The existing CMS search at this center-of-mass energy~\cite{2019398} is based on a data sample corresponding to an integrated luminosity of $36\,\fbinv$ and results in an upper limit on $\mathcal{B}(\Hb\to\ab\ab\to \mmbb)$ of $(1$--$7)\times 10^{-4}$, considering \ma between 20 and 62.5\GeV.
At 8\TeV, the CMS experiment has provided an upper bound of $\mathcal{B}(\Hb\to\ab\ab\to \mmbb)< (2$--$8)\times 10^{-4}$~\cite{Khachatryan:2017mnf}.
In the \ttbb final state, an upper limit of $\mathcal{B}(\Hb\to\ab\ab\to\ttbb)<(3$--$12)\times 10^{-2}$ was reported by CMS using a 36\fbinv dataset at $\sqrt{s}=13\TeV$, where \ma ranged between 15 and 60\GeV~\cite{2016bbtautau}.
The analysis examined both leptonically and hadronically decaying \PGt leptons, the latter denoted by \tauh.

This paper reports an extension of CMS searches~\cite{2019398, 2016bbtautau} with the proton-proton ($\Pp\Pp$) collision data collected in Run~2, corresponding to an integrated luminosity of $\il$ at 13 \TeV.
Improved techniques in these analyses bring higher sensitivity to the allowed branching fractions.
In the \mmbb final state, in particular, a more in-depth study of the signal achieves a greater gain in sensitivity than that offered by the additional LHC data alone.
This channel looks for a bump over the dimuon mass spectrum after a cut-based event selection.
A neural network approach to optimize the signal region (SR) selection provides better sensitivity to the signal processes in the \ttbb channel.
The results in the two final states are combined, and interpretations are provided in different types of 2HDM+S models.

The paper is organized as follows: Sections 2 and 3 discuss the CMS detector and the simulated data samples used in these analyses. The event reconstruction and event selection procedures are presented in Sections~4 and 5, respectively. The background prediction methods are described in Section~6.  Section~7 presents the signal extraction methods, and the discussion of the systematic uncertainties can be found in Section~8. Results and interpretations are detailed in Section~9 and a summary is presented in Section~10. Tabulated results of this analysis are provided in the HEPData record~\cite{hepdata}.
\section{The CMS experiment}
The central feature of the CMS apparatus is a superconducting solenoid of 6\unit{m} internal diameter, providing a magnetic field of 3.8\unit{T}. Within the solenoid volume are a silicon pixel and strip tracker, a lead tungstate crystal electromagnetic calorimeter (ECAL), and a brass and scintillator hadron calorimeter (HCAL), each composed of a barrel and two endcap sections. Forward calorimeters extend the pseudorapidity ($\eta$) coverage provided by the barrel and endcap detectors. Muons are detected in gas-ionization chambers embedded in the steel flux-return yoke outside the solenoid.
The collision data are recorded with the help of Level-1 (L1) trigger, high-level trigger (HLT), and data acquisition systems ensuring high efficiency in selecting interesting physics events~\cite{CMS:2016ngn}.
A more detailed description of the CMS detector, along with a definition of the coordinate system used and the relevant kinematic variables, can be found in Ref.~\cite{Chatrchyan:2008aa}.
\section{Simulated event samples}
\label{sim}
Simulated samples are used to design and optimize the analysis strategy and, where needed, to estimate background contributions.
A number of Monte Carlo (MC) event generators are used to produce events using either leading order (LO) or NLO matrix element calculations.
In all cases, parton showering and fragmentation are implemented using \PYTHIA (version 8.212)~\cite{Sjostrand:2014zea}.
The description of parton distribution functions (PDFs) relies on the NNPDF3.1 set~\cite{Ball:2017nwa}.
Jets produced at the matrix element level are matched with those generated by \PYTHIA using the MLM~\cite{Alwall:2008qv, Alwall:2007fs} method for LO samples.
The FxFx matching~\cite{Frederix:2012ps} is implemented in the case of NLO samples generated with \MGvATNLO (version 2.2.2 for the simulation of the 2016 data and 2.4.2 for 2017--2018)~\cite{Alwall:2014hca}.
For the underlying event description, the CUETP8M1~\cite{Skands:2014pea} tune was used for MC samples simulating the 2016 data, while for those simulating the 2017--2018 data, the CP5~\cite{CP5tune:2020} tune was employed.
The \GEANTfour~\cite{GEANT4:2002zbu, Allison:2006ve} package has been used for the detector simulation.
To model the effect of additional collisions within the same or adjacent bunch crossings (pileup), minimum bias interactions are simulated and superimposed on the hard-scattering events.
Simulated events are then reweighted to reproduce the pileup distribution in data.

The $\Hb\to\ab\ab\to\mmbb$ signal events are generated with the NMSSMHET model~\cite{PhysRevD.90.075004} using \MGvATNLO (version 2.6.5) at LO~\cite{Alwall:2007fs}.
Both ${\Pg\Pg}$F and VBF Higgs boson production mechanisms are considered, within the \ab mass range of 15--60\GeV.
While the ${\Pg\Pg}$F samples are generated with 5\GeV steps in $\ma$, the VBF samples are generated only for $\ma$ of 20, 40, and 60\GeV.
Interpolation methods are used to estimate the signal yield and the shape of the dimuon resonance for all mass hypotheses.
Similar settings are used to produce $\Hb\to\ab\ab\to\ttbb$ signal events at 11 \ab masses between 12 and 60\GeV, for both ${\Pg\Pg}$F and VBF Higgs boson production modes.

The major backgrounds for the analyses are the Drell--Yan (DY) process (${\PZ}/\gamma^*$+jets), the production of a top quark-antiquark pair with additional jets (denoted {\toppair}+jets), single top quark production, and massive vector boson pair production (Diboson).
In the \mmbb channel, the background estimation is performed using methods fully based on control samples in data with no reference to simulation.
Simulated background samples are, however, used to optimize the signal selection criteria.
In the \ttbb channel, only the backgrounds from DY production with $\PZ\to\PGt\PGt$, QCD multijet events in the \emu final state, and events with jets misidentified as $\tauh$ candidates are estimated using control samples in data.

The DY process in the dilepton final state is modeled using \MGvATNLO.
Based on the dilepton invariant mass (\mellell) threshold at the generator level, two DY samples are considered, one with $\mellell>50\GeV$ and the other with $10<\mellell<50\GeV$.
The high-mass DY samples are produced at (N)LO, with up to four (two) additional partons at the matrix element level.
For the low mass, samples are primarily produced at LO with additional partons, similar to those of high mass, while NLO and LO samples inclusive in number of jets are also utilized.
In the \mmbb analysis, the NLO samples at high mass are employed, and at low mass, NLO QCD K-factors are applied to the LO cross section.
An uncertainty of 30\% is considered on these K-factor corrections, as they are extracted from NLO low-mass samples with limited number of events.
The accuracy of the DY sample is found to be sufficient for optimization purposes, which is the only use of the simulated backgrounds in the \mmbb channel.
The \ttbb analysis makes use of LO DY samples in the entire mass range.
The cross sections are normalized to next-to-NLO (NNLO) in QCD using K-factors~\cite{CMS:2022kdi}.
In addition, the $\PZ$ boson \pt distribution is corrected by reweighting simulated events to data in bins of \mellell and the \pt of the dilepton system.

The \POWHEG \textsc{box} v2.0 framework~\cite{Nason:2004rx, Frixione:2007vw, Alioli:2010xd, Alioli:2010xa} event generator is used to produce {\toppair}+jets and single top events at NLO.
The simulated {\toppair}+jets events are reweighted to match the top quark \pt distribution at NNLO QCD and NLO EW~\cite{Czakon_2017} precision.
Diboson and \wjets\ events are generated by \MGvATNLO.
Similar to the high-mass DY sample, \wjets\ events are simulated with up to four additional partons at the matrix element level for all years.
The {\toppair}+jets, DY, and \wjets\ samples are normalized to cross section values accurate to NNLO in QCD~\cite{Czakon:2011xx, Botje:2011sn, Martin:2009bu, Gao:2013xoa, Ball:2012cx, Campbell:2020fhf, PDF4LHCWorkingGroup:2022cjn, Melnikov:2006kv, Martin:2009iq}.
All SM backgrounds containing the Higgs boson are generated using \POWHEG v2.0 at NLO~\cite{Alioli:2008tz, Bagnaschi:2011tu, Nason:2009ai, Luisoni:2013cuh, Hartanto:2015uka}.
\section{Object reconstruction}
\label{sec:obj}
The \mmbb and \ttbb analyses together reconstruct a diverse set of final-state particles for a $\Hb\to\ab\ab$ signal. The \mmbb analysis relies on the presence of two prompt muons. In the \ttbb channel, on the other hand, final states with at least one \PGt lepton decaying to an electron or muon, \ie, \emu, \etau, and \mutau, are considered. The \PGt lepton decays resulting in same-flavor leptons, or in two \tauh candidates, are not included as they bring negligible sensitivity to the analysis. The signal acceptance of $\tauh\tauh$ is very low due to high trigger thresholds, whereas $\Pe\Pe$ and $\PGm\PGm$ final states suffer from low branching fractions.

The particle-flow (PF) algorithm~\cite{Sirunyan:2017ulk} is used to reconstruct and identify each individual particle (PF candidate) in the event, with an optimized combination of information from the various elements of the CMS detector. The energy of photons is measured in ECAL. The energy of electrons is determined from a combination of the electron momentum at the primary interaction vertex as measured by the tracker, the energy of the corresponding ECAL cluster, and the energy sum of all bremsstrahlung photons spatially compatible with originating from the electron track. The energy of muons is obtained from the curvature of the corresponding track. The energy of charged hadrons is evaluated via a combination of their momentum measured in the tracker and the matching of the ECAL and HCAL energy deposits, corrected for the response function of the calorimeters to hadronic showers. Finally, the energy of neutral hadrons is obtained from the corresponding corrected ECAL and HCAL energies.

The primary vertex (PV) is taken to be the vertex corresponding to the hardest scattering in the event, identified using the tracking information alone, as described in Section 9.4.1 of Ref.~\cite{CMS-TDR-15-02}.

Muons can be produced directly in \ab decays in the \mmbb final state, or from decays of the \PGt leptons in the \ttbb channel. In both analyses, muons must lie within $\abs{\eta} < 2.4$.
The \pt threshold for muons in the \ttbb analysis depends on the trigger selection, see Section~\ref{sec:evt} and Table~\ref{tab:triggers}.
In the \mutau final state, it is required to be 1\GeV greater than the HLT muon \pt threshold in order to be in a region where the efficiency of the respective trigger is independent of \pt.
The muon \pt requirement in the \emu final state, selected with an \emu trigger, is 24\,(13)\GeV when the muon is the leading (subleading) lepton in the pair.
In the \mmbb analysis, the leading (subleading) muon \pt must exceed 17\,(15)\GeV.
The two muons are required to have an opposite electric charge and to be separated by a minimum $\Delta R \equiv \sqrt{(\Delta\eta)^{2} + (\Delta\phi)^{2}} = 0.4$, where $\phi$ is the azimuthal angle of the particle's momentum in the plane perpendicular to the beam line.
In cases where more than two muons satisfy these criteria, the pair with the highest \pt are considered.

In order to suppress contributions from nonprompt decays of hadrons and from their shower penetration in the muon detectors, selected muons must pass dedicated identification requirements.
The so-called tight identification~\cite{Sirunyan:2018fpa} is used in the \mmbb analysis with an efficiency varying between 95 and 99\%, depending on $\eta$, where the data and simulation agree within 1--3\%.
Looser requirements for muons, known as medium identification criteria~\cite{Sirunyan:2018fpa}, are employed in the \ttbb analysis, with an overall efficiency of 99.5\% for simulated \PW and \PZ events.

The lepton isolation variable $\irel$ is calculated by summing the transverse energy deposited by other particles in a cone of size $\Delta R$ = 0.4\,(0.3) around the muon (electron) and dividing by the lepton \pt.
The contribution of charged particles from pileup is suppressed by requiring the charged particles to be associated with the PV.
An average pileup energy is subtracted from the total energy of neutral particles and photons within the isolation cone, since vertex association is not known in this case.
Muons are required to pass $\irel<0.15$ in the \ttbb analysis.

In the \mmbb analysis, a looser requirement of $\irel<0.25$ is imposed, which results in about 99\% efficiency for muons with $\pt>20\GeV$, independent of $\eta$~\cite{Sirunyan:2018fpa}.
Electrons from \PGt lepton decays are selected within $\abs{\eta}<2.4$ with different \pt thresholds, according to the \ttbb final state. In the \emu channel, the threshold is 24\GeV if the electron is the leading lepton.
Otherwise, it is reduced to 13\GeV.
In the \etau channel, the electron \pt must be more than 1\GeV beyond the HLT threshold.
A multivariate analysis (MVA) discriminant is used to identify electrons.
The MVA exploits several properties of the electron candidate, including energy deposits in the ECAL, the quality of the associated track, and the shower shape in the calorimeters~\cite{Electrons:2021}.
The chosen MVA working point has a 90\% efficiency to correctly identify an electron.
Identified electrons are further required to be isolated, fulfilling $\irel<0.10$. In the \emu channel of the \ttbb analysis, the electron must be separated from the muon by $\Delta R \geq 0.3$ and have an opposite electric charge.
For both electrons and muons, correction factors for the reconstruction and identification efficiencies are obtained from data and applied to simulation.

Jets are reconstructed by clustering the charged and neutral PF candidates using the anti-$\kt$ algorithm~\cite{Cacciari:2008gp, Cacciari:2011ma} with a distance parameter of 0.4, up to $\abs{\eta} < 4.7$ for tagging VBF events.
The reconstructed jet energy is corrected for effects from the detector response as a function of the jet $\pt$ and $\eta$.
Furthermore, contamination from pileup and electronic noise is subtracted using the charged-hadron subtraction method~\cite{Sirunyan:2017ulk}.
To achieve a better agreement between data and simulation, an extra $\eta$-dependent smearing is performed on the jet energy in simulated events~\cite{CMS-PAS-JME-16-003,CMSJetPaper}.
Events are required to have at least two (one) jets with $\abs{\eta}<2.4$ and $p_{\rm T}>15\,(20)$\GeV in the \mmbb (\ttbb) analysis.
Jets are required to be separated from any selected electron, muon, or \tauh by $\Delta R > 0.4\,(0.5)$ in the \mmbb (\ttbb) analysis.

Both channels rely on identifying jets that likely originate from \PQb quarks.
The \textsc{DeepJet} flavor classification algorithm~\cite{Bols:2020bkb,Sirunyan:2017ezt} is used to tag \PQb jets.
Three different working points on the {\PQb} tagging discriminator values correspond to 0.1, 1, and 10\% misidentification probabilities, known respectively as tight (T), medium (M), and loose (L) working points.
The misidentification probability to tag a light-flavour jet as a \PQb jet is measured in inclusive QCD multijet MC samples, and they depend on the $\pt$ and $\eta$ of the jet.
The corresponding \PQb tagging efficiencies, measured in {\toppair}+jets events, are about 65, 80, and 95\%, respectively~\cite{CMS-DP-2018-058}.
In the \mmbb analysis, the selected jet with the higher \PQb tagging score is required to pass the tight working point whereas the second one fulfills the loose \PQb tagging requirements.  In this paper the latter is referred to as the 'looser' \PQb jet.
In the \ttbb analysis, the medium working point is used to identify \PQb jets.
The shape of the distribution of the \PQb tagging discriminator, and thus the \PQb tagging efficiencies, can be different between data and simulation.
Since the \mmbb analysis relies on the \PQb tagging discriminator distribution, shape-based corrections are applied on simulation to match the data.
A similar method is used in the \ttbb final state, which, by construction corrects the \PQb tagging efficiency for all \PQb tagging discriminator scores.

The hadron-plus-strips algorithm~\cite{CMS:2018jrd} with anti-\kt jets as seeds is used to reconstruct the hadronically decaying \PGt leptons.
The algorithm combines one or three tracks with energy deposits in the calorimeters to identify the \PGt lepton decay modes.
Neutral pions are reconstructed as strips with a dynamic size in $\eta$-$\phi$ from reconstructed electrons and photons, where the strip size varies as a function of the $\pt$s of the electron or photon candidate. The \pt of the \tauh candidates are required to be 5\GeV greater than the threshold at the trigger level. In events triggered by single leptons, the \tauh \pt must exceed 20\GeV. The pseudorapidity of the \tauh candidate also depends on the trigger. It is restricted to $\abs{\eta} < 2.1$ if a \tauh identification is performed at the HLT, and to $\abs{\eta} < 2.3$ otherwise.
To distinguish genuine \tauh decays from electrons, muons, or jets originating from the hadronization of quarks or gluons, the \textsc{DeepTau} algorithm~\cite{CMS:2022prd} is used.
Information from all individual reconstructed particles near the \tauh candidate axis is combined with properties of the \tauh candidate.
The probability for a jet to be misidentified as a \tauh candidate by the \textsc{DeepTau} algorithm depends on the \pt and the jet flavor.
In simulated {\PW}+jets events,  the misidentification rate for jets is estimated to be 0.4\% for a genuine \tauh identification efficiency of 70\%.
The misidentification rate for electrons (muons) is 2.6\% (0.03\%) for a genuine \tauh identification efficiency of 80\% ($>99$\%).
In the \etau and \mutau final states of the \ttbb channel, the \tauh candidate must be separated from the electron or muon by $\Delta R \geq 0.4$ and they must be oppositely charged.

The missing transverse momentum vector \ptvecmiss is computed as the negative vector sum of the transverse momenta of all the PF candidates in an event, and its magnitude is denoted as \ptmiss~\cite{CMS:2019ctu}. The \ptvecmiss is modified to account for corrections to the energy scale of the reconstructed jets in the event.
Anomalous high-\ptmiss events can be due to a variety of reconstruction failures, detector malfunctions or noncollision backgrounds. Such events are rejected by event filters that are designed to identify more than 85--90\% of the spurious high-\ptmiss events with a mistagging rate less than 0.1\%~\cite{CMS:2019ctu}.
\section{Event selection}
\label{sec:evt}
Table~\ref{tab:triggers} summarizes the different \pt criteria for online reconstructed electrons, muons and $\tauh$s in the \mmbb and \ttbb channels.

\begin{table*}[ht!]
\centering
\topcaption{The electron, muon, and \tauh \pt thresholds in\GeV at trigger level for the \mmbb and \ttbb channels.}
\begin{tabular}{lcccccccc}
\hline
\multirow{3}{*}{ Year } & \multirow{3}{*}{ Single/dilepton trigger \pt } & \mmbb & \multicolumn{6}{c}{\ttbb} \\
& & & \multicolumn{2}{c}{\emu} & \multicolumn{2}{c}{\etau} & \multicolumn{2}{c}{\mutau}  \\
& & $\mu$ & \Pe & $\mu$ & \Pe & \tauh & $\mu$ & \tauh \\
\hline
\multirow{3}{*}{ 2016 } & Single lepton      &  24     &        \NA       &       \NA      &         25          &          \NA         &          22         &         \NA    \\
                        & \pt-leading lepton        &  17     &        23       &       23      & \multirow{2}{*}{\NA} & \multirow{2}{*}{\NA} &          \NA         &         20    \\
                        & \pt-subleading lepton     &  8      &        12       &       8       &                     &                     &          19         &         \NA    \\
\hline
\multirow{3}{*}{ 2017 } & Single lepton       & 24      &        \NA       &       \NA      &       27, 32        &          \NA         &       24, 27        &         \NA    \\
                        & \pt-leading lepton        &  17     &        23       &       23      &         \NA          &          30         &          \NA         &         27    \\
                        & \pt-subleading lepton     &  8      &        12       &       8       &         24          &          \NA         &          20         &         \NA    \\
\hline
\multirow{3}{*}{ 2018 } & Single lepton      & 24      &        \NA       &       \NA      &       32, 35        &          \NA         &       24, 27        &         \NA    \\
                        & \pt-leading lepton        &  17     &        23       &       23      &         \NA          &          30         &          \NA         &         27    \\
                        & \pt-subleading lepton     &  8      &        12       &       8       &         24          &          \NA         &          20         &         \NA    \\
\hline
\end{tabular}
\label{tab:triggers}
\end{table*}

The $\mmbb$ event candidates are selected based on the requirement that either one or both muons are reconstructed at the HLT.
Passing the double-muon trigger necessitates two isolated muons with $\pt$ exceeding thresholds of 17 and 8\GeV, which increases to 24\GeV for an isolated muon in the single-muon trigger path.
Accepting events from both single- and double-muon triggers improves the trigger efficiency by including events in which the second muon is not reconstructed at the trigger level.

Depending on the decay of the {\PGt} lepton and the data-taking period, the \ttbb candidates must pass either a single-electron, single-muon, $\Pe\PGm$, $\Pe\tauh$, or $\PGm\tauh$ trigger selection.
The single-muon, $\Pe\PGm$ and $\PGm\tauh$ triggers require the reconstructed muon to be isolated, while electron isolation is required for the single-electron, $\Pe\PGm$ and $\Pe\tauh$ triggers.
Two $\Pe\PGm$ dilepton triggers have been used for all data-taking years, having \pt thresholds of 23\,(23) and 12\,(8)\GeV for the \pt-leading and -subleading lepton of the trigger in the case of electrons (muons).
The single-muon and single-electron triggers with \pt thresholds of 22 and 25\GeV, respectively, are used for analyzing the 2016 data.
The \pt thresholds of electron and \tauh candidates are, respectively, 24 and 30\GeV for the $\Pe\tauh$ dilepton trigger in the 2017-2018 data.
For the $\PGm\tauh$ dilepton trigger, the \pt thresholds of muon and \tauh are, respectively, 19\,(20) and 20\,(27)\GeV for the data taken during 2016 (2017-2018).
The increase in the \pt threshold is necessary to control the trigger rate at a larger instantaneous luminosity.
Similarly, the \pt requirements are tightened for single-lepton triggers across the years.
This results in two different thresholds for single-electron and muon triggers for 2017-2018.

\begin{figure*}[!ht]
        \centering
        \includegraphics[width=0.42\textwidth]{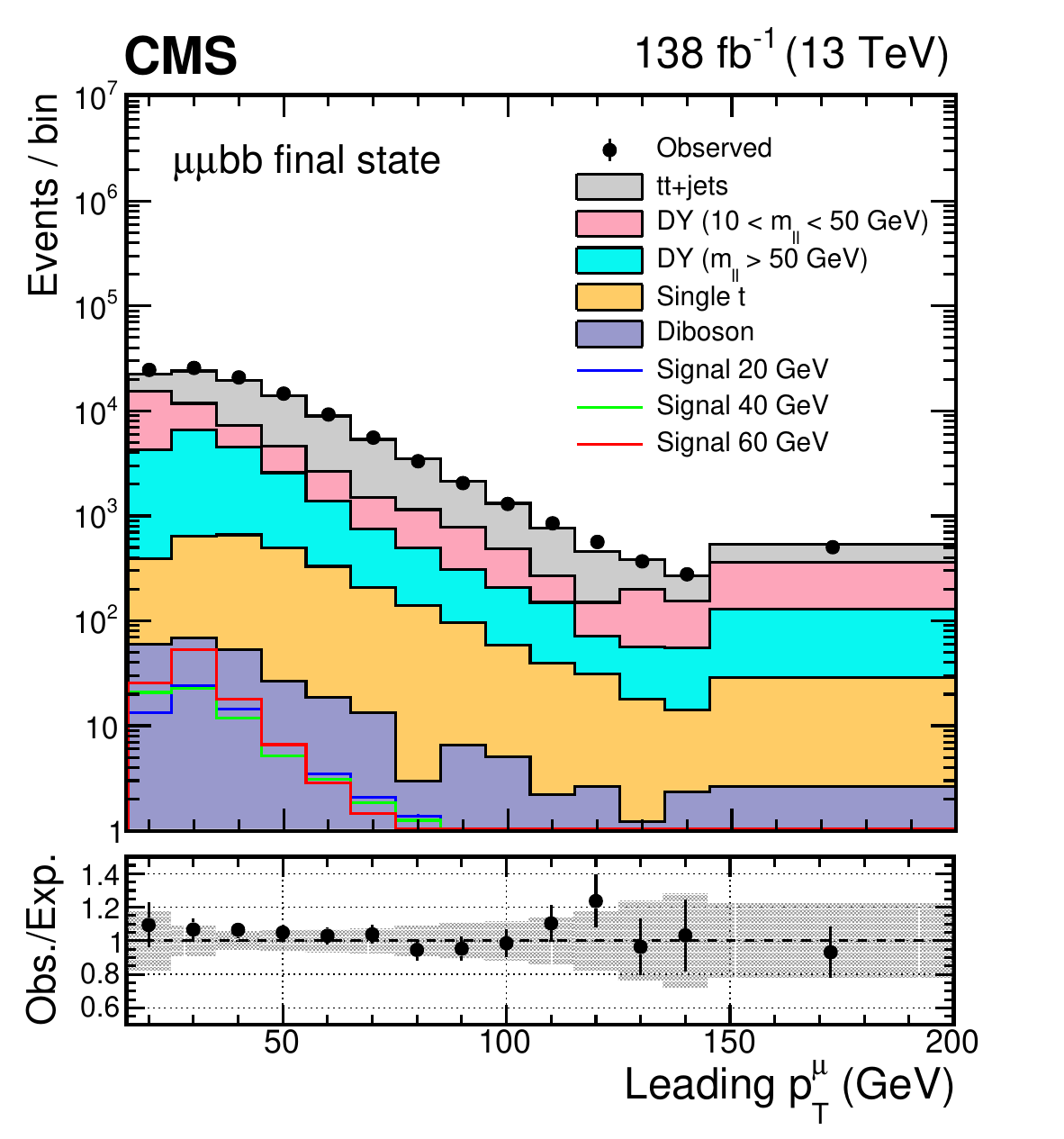}
        \includegraphics[width=0.42\textwidth]{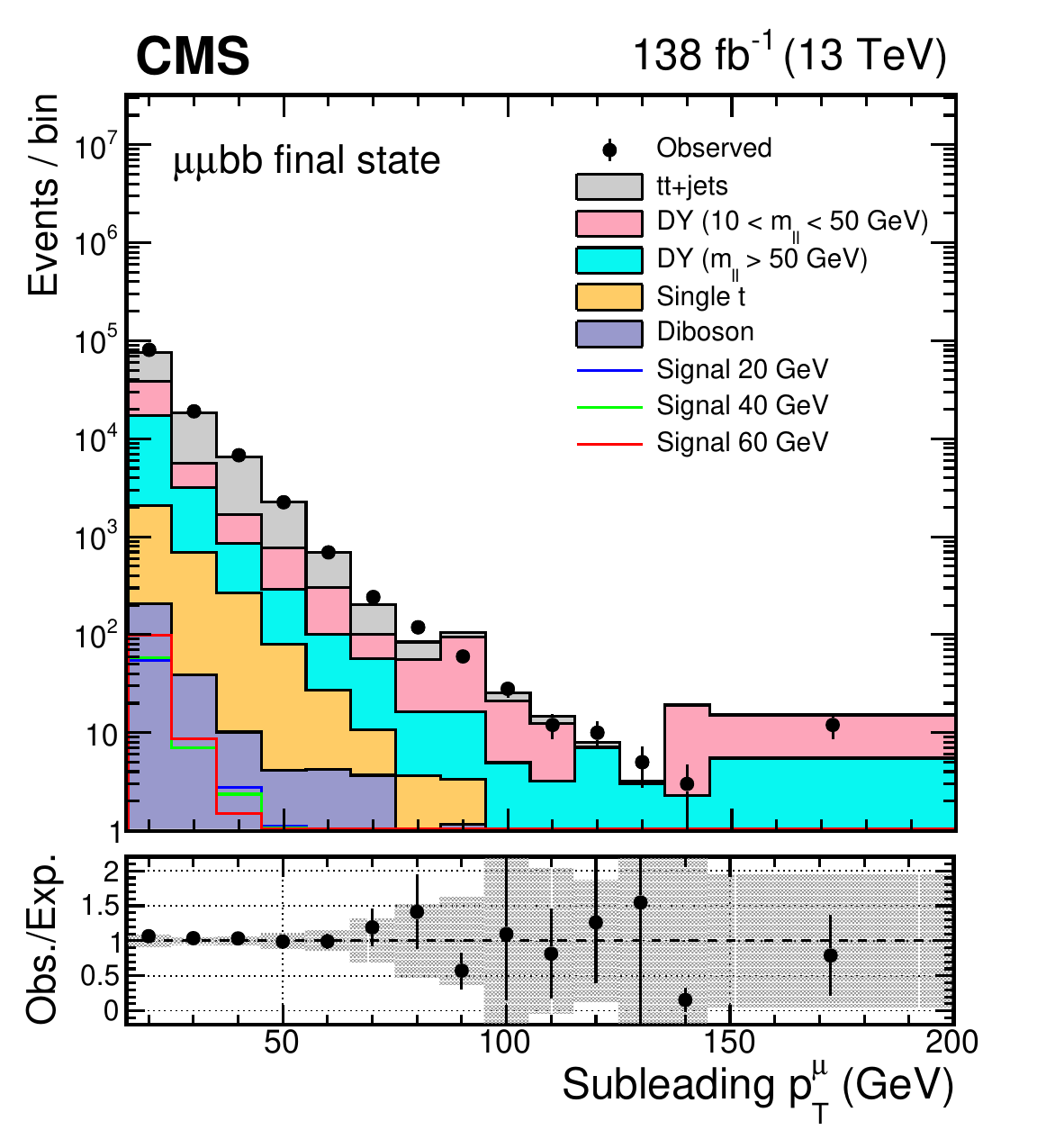} \\
        \includegraphics[width=0.42\textwidth]{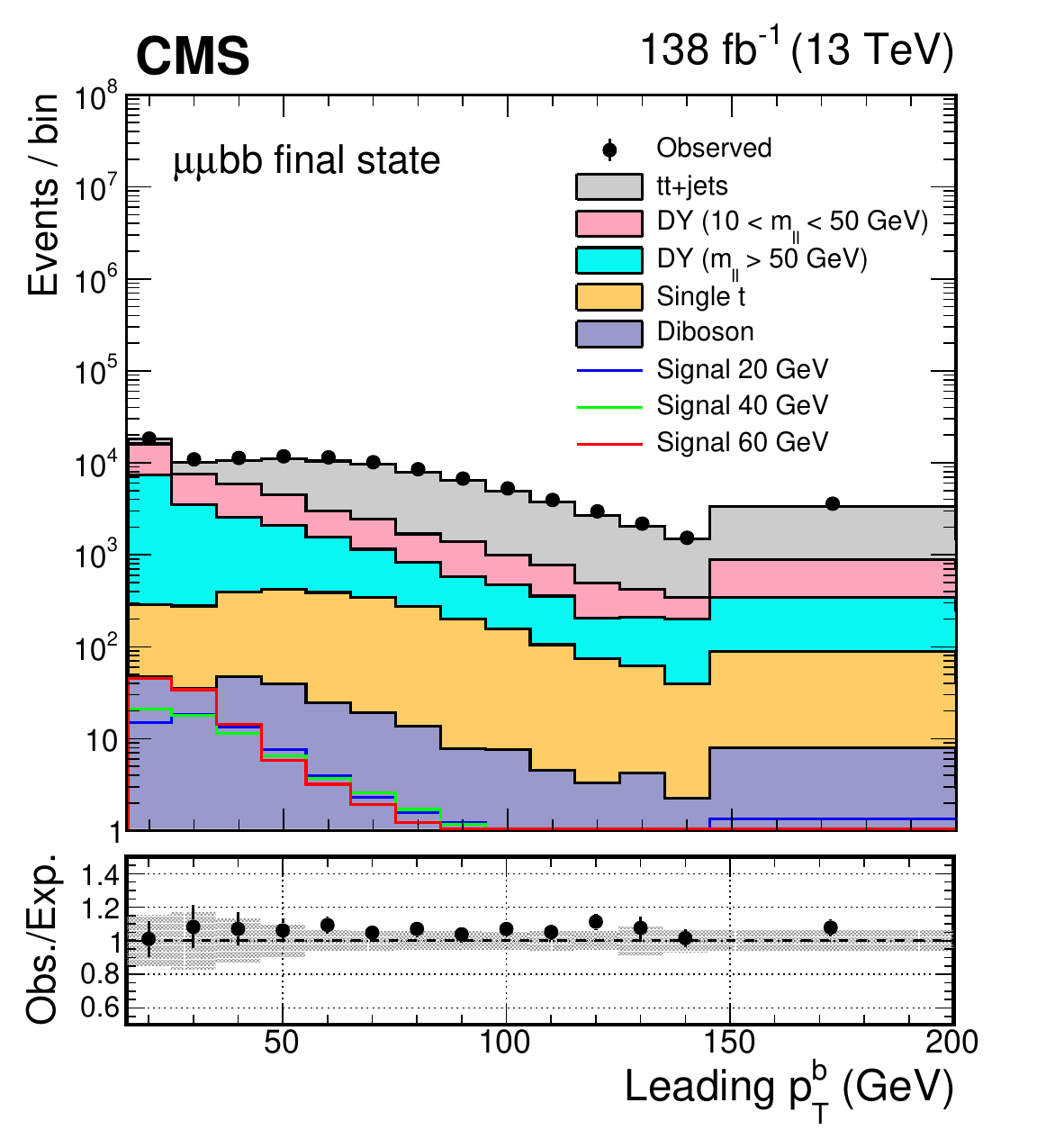}
        \includegraphics[width=0.42\textwidth]{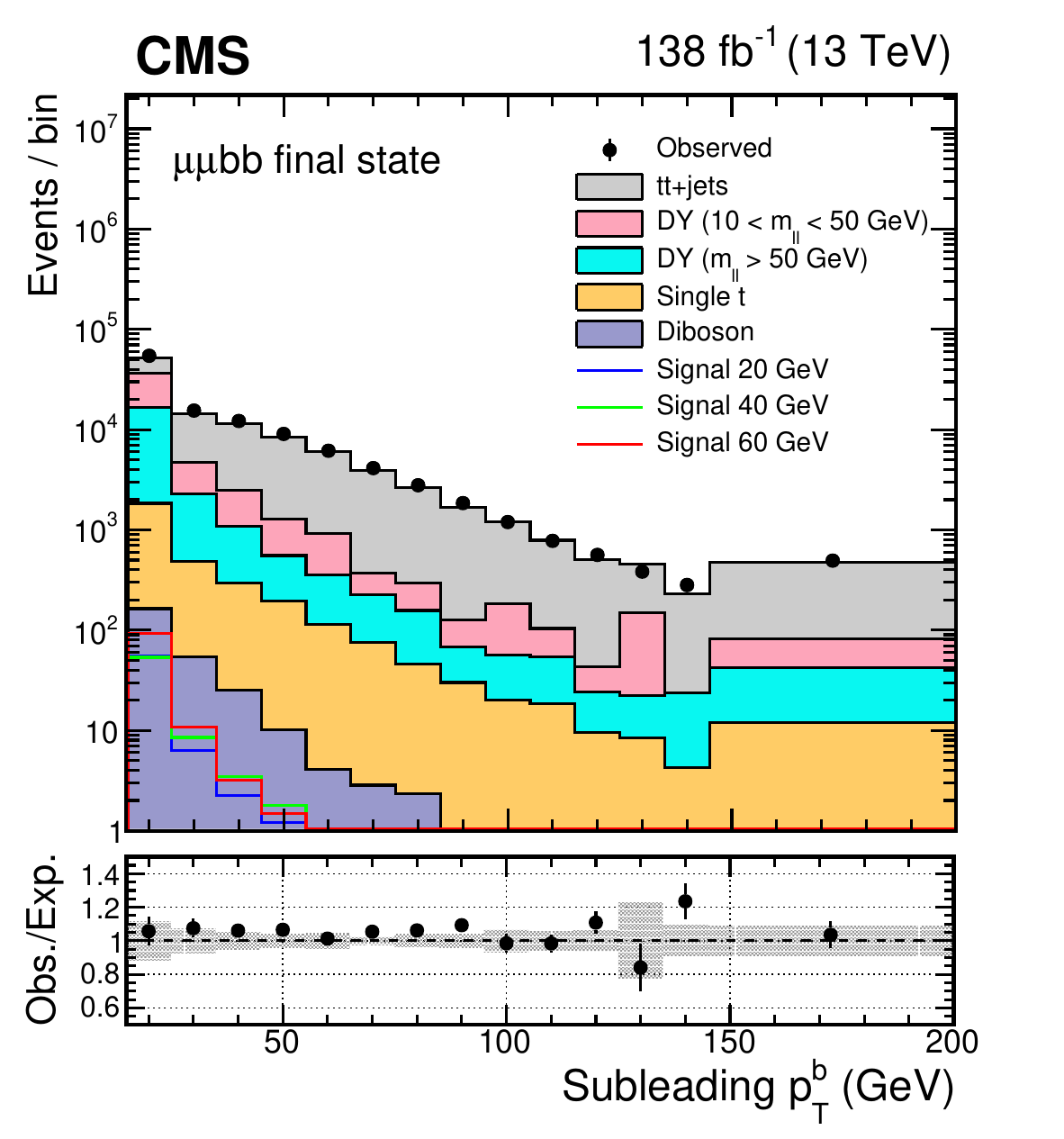}
        \caption{The distributions of leading and
          subleading (upper) muon $\pt$ and (lower) \PQb jet $\pt$ in the selected events. The
          uncertainty band in the lower panel represents the
          limited size of the simulated samples together with a
          30\% uncertainty in the low-mass DY cross
          section. Simulated samples are normalized
          using the corresponding theoretical cross
          sections. To evaluate the normalization of the signal, SM Higgs boson cross sections are multiplied by the $\mathcal{B}(\ab\ab\to\mmbb)$ value that is calculated in the Type III model with $\tanb = 2$.}
                \label{fig:presel1}
\end{figure*}

\begin{figure}[!ht]
        \centering
        \includegraphics[width=0.42\textwidth]{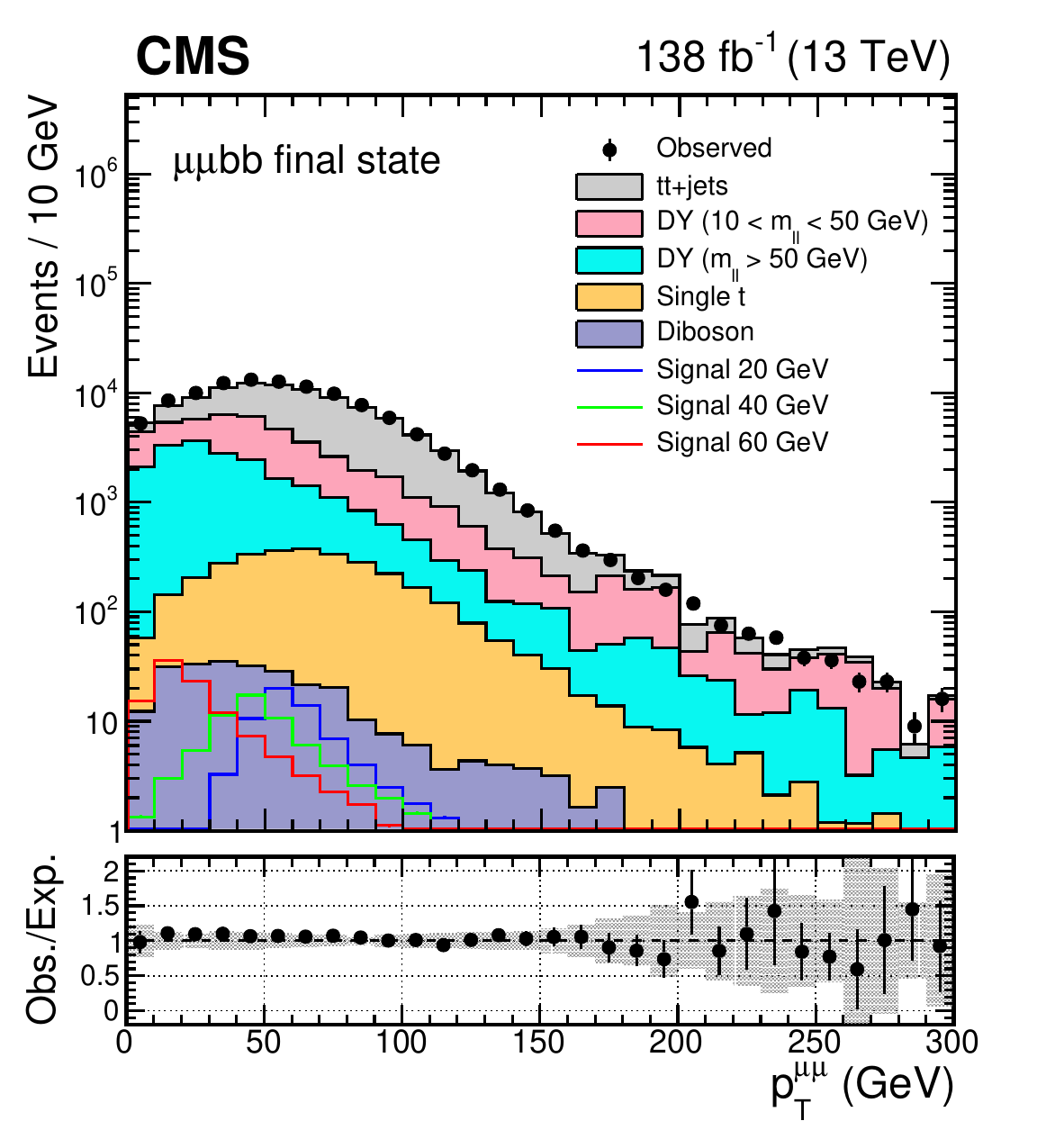}
        \includegraphics[width=0.42\textwidth]{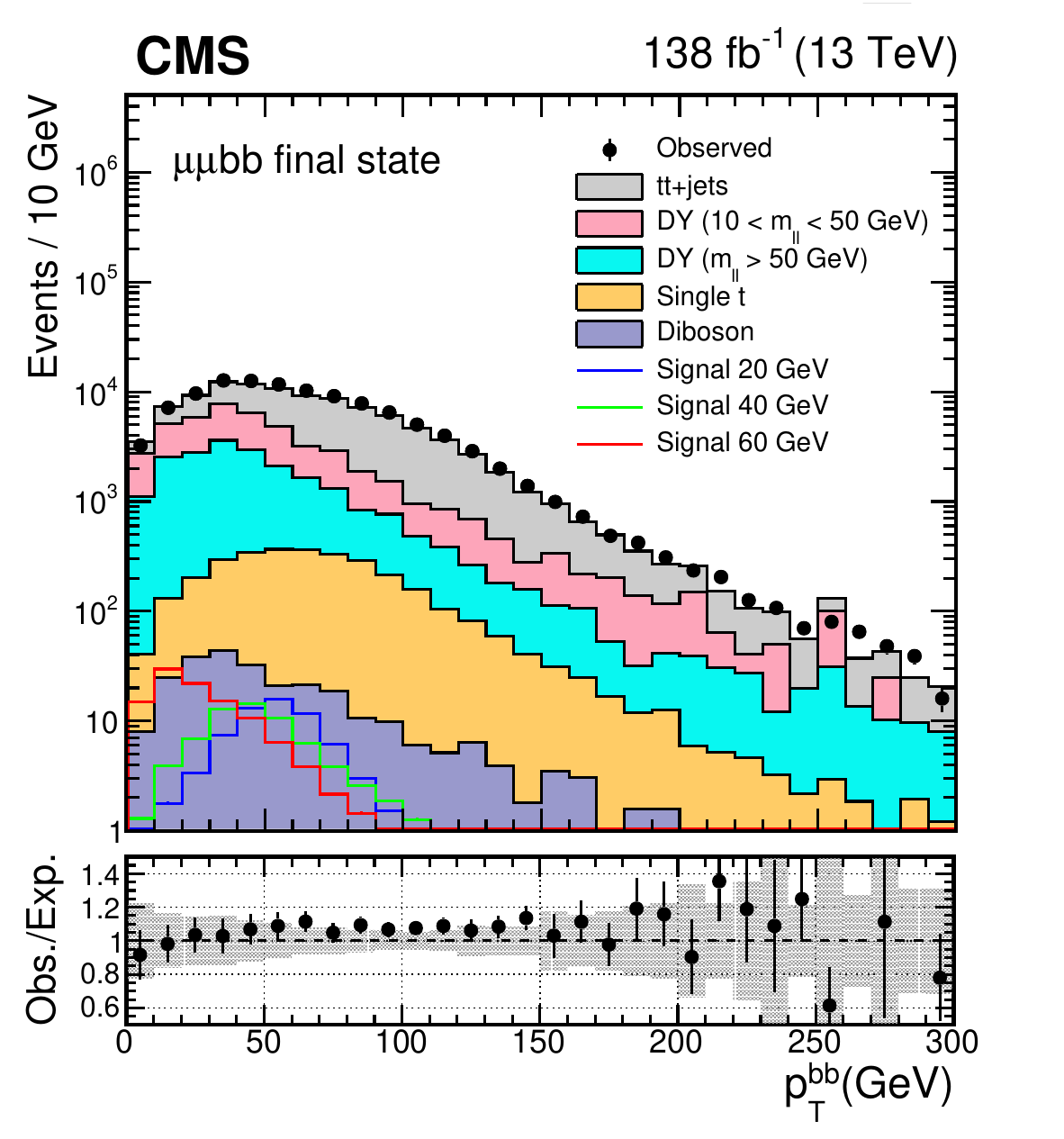}
        \caption{The \pt distributions of the (\cmsLeft) dimuon systems and
          (\cmsRight) di-\PQb-jet system. The uncertainty band in the
          lower panel represents the limited size of the simulated samples
          together with a 30\% uncertainty in the low-mass DY
          cross section. Simulated samples are normalized to
          using the corresponding theoretical cross sections. To evaluate the normalization of the signal, SM Higgs boson cross sections are multiplied by the $\mathcal{B}(\ab\ab\to\mmbb)$ value that is calculated in the Type III model with $\tanb = 2$.}
        \label{fig:presel2}
\end{figure}

Offline, in the \mmbb channel events are required to have two muons and at least two \PQb jets passing the kinematic, identification, and isolation criteria detailed in Section~\ref{sec:obj}. While the final search in this channel is performed for \ma between 15 and 62.5\GeV, events are selected with a dimuon invariant mass, \mmumu, between 14 and 70\GeV. The additional sidebands in \mmumu help model the backgrounds at the boundaries. To reduce the background contribution from {\toppair}+jets, events with $\ptmiss > 60\GeV$ are rejected.
The selection yields a total of 109\,821 data events while the corresponding expected yield from simulated backgrounds is $103\,900 \pm 7300$.
The background contribution should be compared with about 80--100 expected signal events, depending on \ma, from both ${\Pg\Pg}$F and VBF Higgs boson production. The branching fraction $\mathcal{B}(\ab\ab\to\mmbb)$ is evaluated in the Type III model with $\tanb=2$.
Figure~\ref{fig:presel1} shows, in data and simulation, the \pt distributions of the \pt-leading and -subleading muons and \PQb jets.
Although the estimation of backgrounds in this analysis does not rely on simulation, the observed level of agreement between data and simulation justifies the use of simulated events to optimize the sensitivity.
Figure~\ref{fig:presel2} shows distributions for the \pt of the dimuon ($\pt^{\PGm\PGm}$) and the di-{\PQb}-jet systems ($\pt^{\PQb\PQb}$).

\begin{figure}[!ht]
        \centering
        \includegraphics[width=0.42\textwidth]{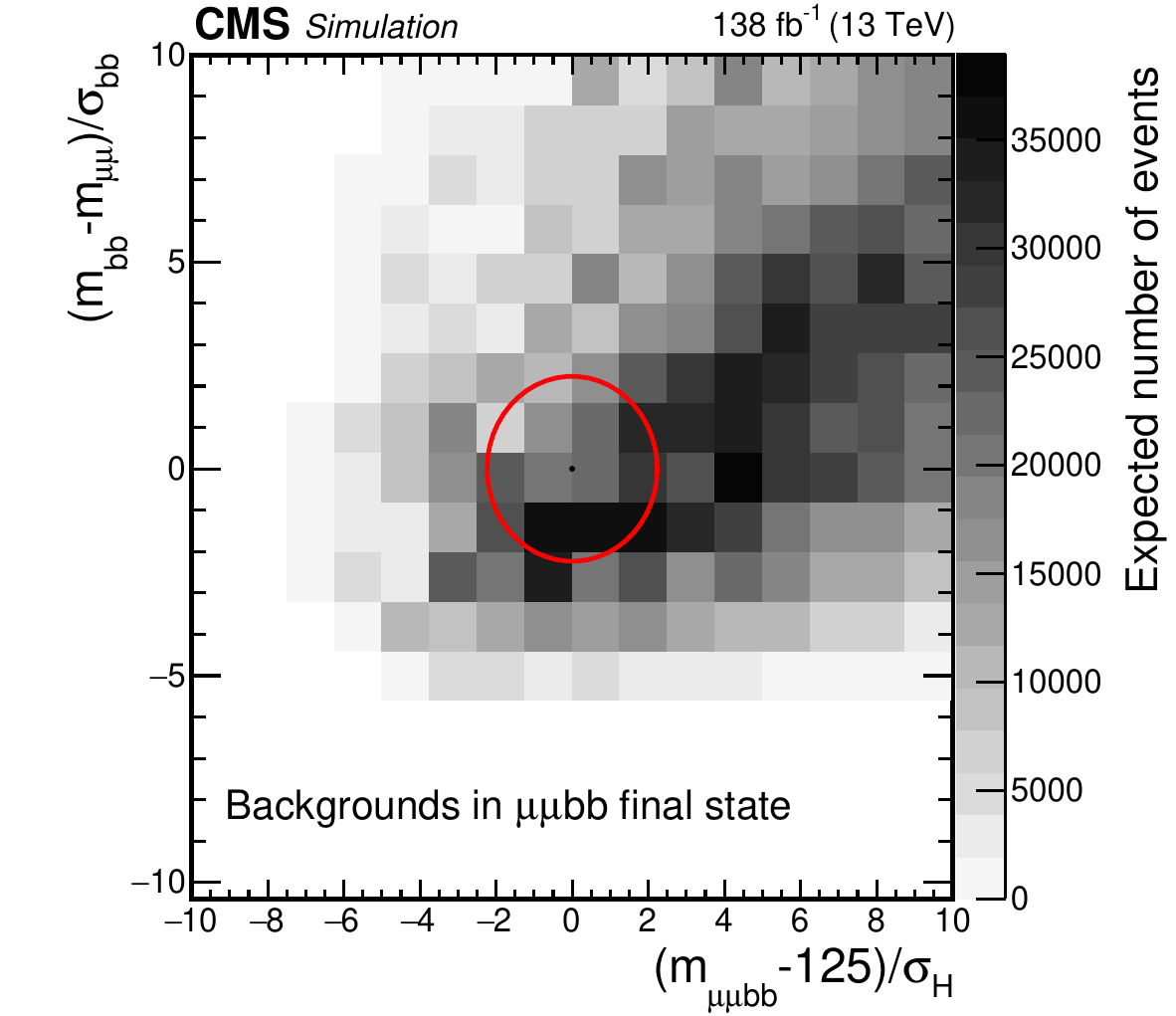}
        \includegraphics[width=0.42\textwidth]{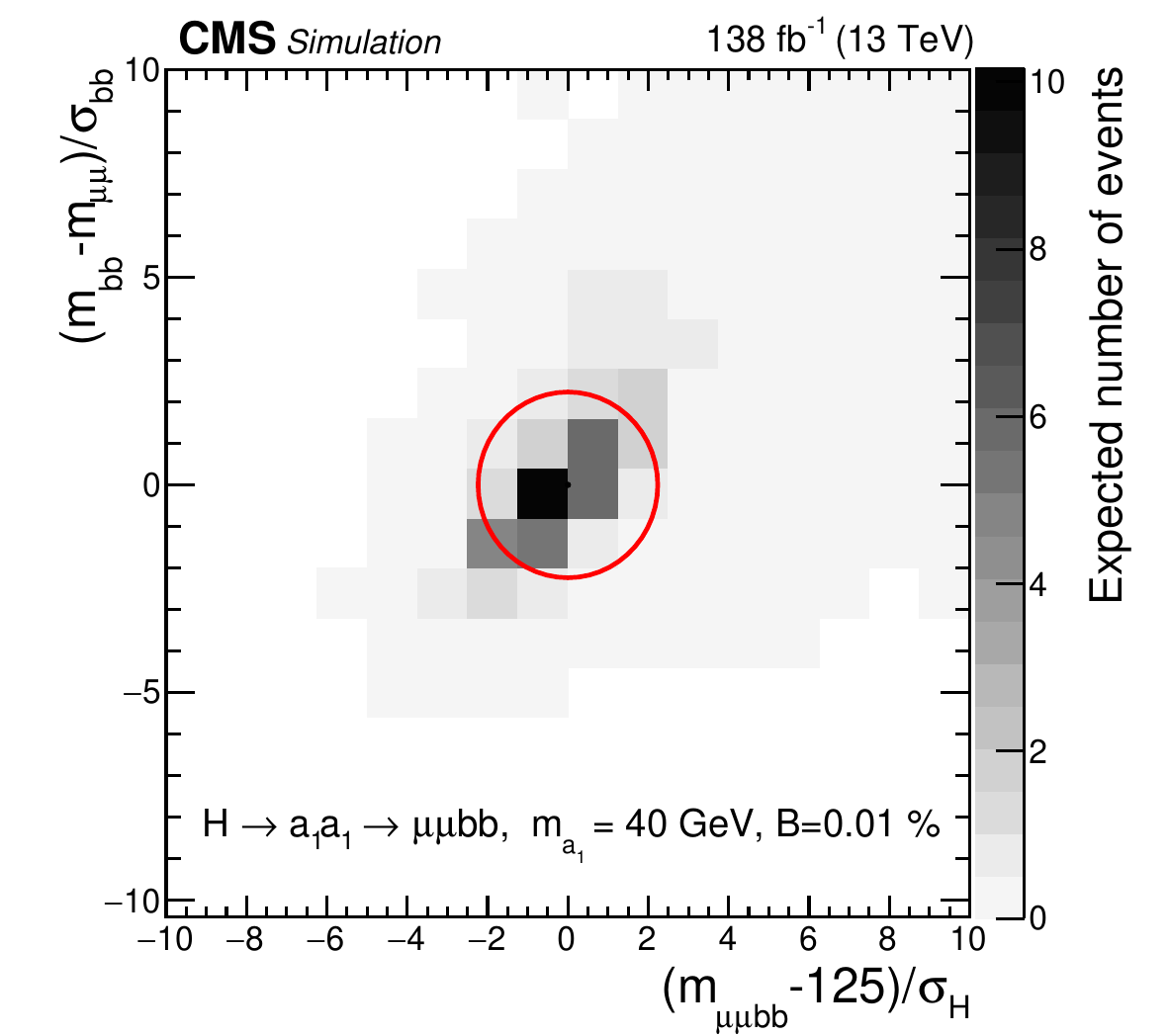}
        \caption{The distribution of \chib versus \chih as defined in
          Eq.~(\ref{eq:chil}) for (\cmsLeft) simulated background
          processes and (\cmsRight) the signal process with $\ma =
          40\GeV$. The contours indicate lines of constant $\chit^2$.
          The gray scale represents the expected yields in data. To evaluate the yield of the signal, SM Higgs boson cross sections are multiplied by the $\mathcal{B}(\ab\ab\to\mmbb)$ value that is calculated in the Type III model with $\tanb=2$.
}
        \label{fig:chi2TD}
\end{figure}

To further suppress backgrounds, a $\chit^2$ variable is defined as $\chit^2 = \chib^2 + \chih^2$. It examines the compatibility of \mmumu and \mbb with \ma, and of \mfour with the Higgs boson mass in signal events. The components of $\chit^2$ are defined as
\begin{equation}
\label{eq:chil}
\chib = \chibl \quad \text{and} \quad \chih = \chihl.
\end{equation}
The variables $\resbb$ and $\resh$ are the mass resolutions of the
di-{\PQb}-jet system and the Higgs boson candidate,
respectively, which are derived from Gaussian fits to simulated distributions of \mbb and the mass of the Higgs boson candidate. While $\resh$ is found to be constant, $\resbb$ increases linearly with \ma and is modeled as
a function of \mmumu ($\resbb = a\mmumu + b$), assuming
$\mmumu=\ma$.  The $\chit^2$
variable is evaluated on an event-by-event basis.  It was shown in Ref.~\cite{2019398} that applying a
threshold on $\chit^2$ leads to a large suppression of backgrounds
while keeping the majority of signal events. Such a requirement
translates to a circle centered at zero in the 2D-plane of \chib and
\chih, as shown in Fig.~\ref{fig:chi2TD}. However, the \chib and \chih components are clearly correlated as can be seen in Fig.~\ref{fig:chi2TD} (\cmsRight). This leads to a loss of signal efficiency when imposing the circular requirement. In addition, both \chib and \chih distributions are slightly biased away from zero, adding more inefficiencies. Therefore, in the current analysis the definitions of the variables were adjusted to be unbiased and uncorrelated.

\begin{figure}[!h]
        \centering
        \includegraphics[width=0.42\textwidth]{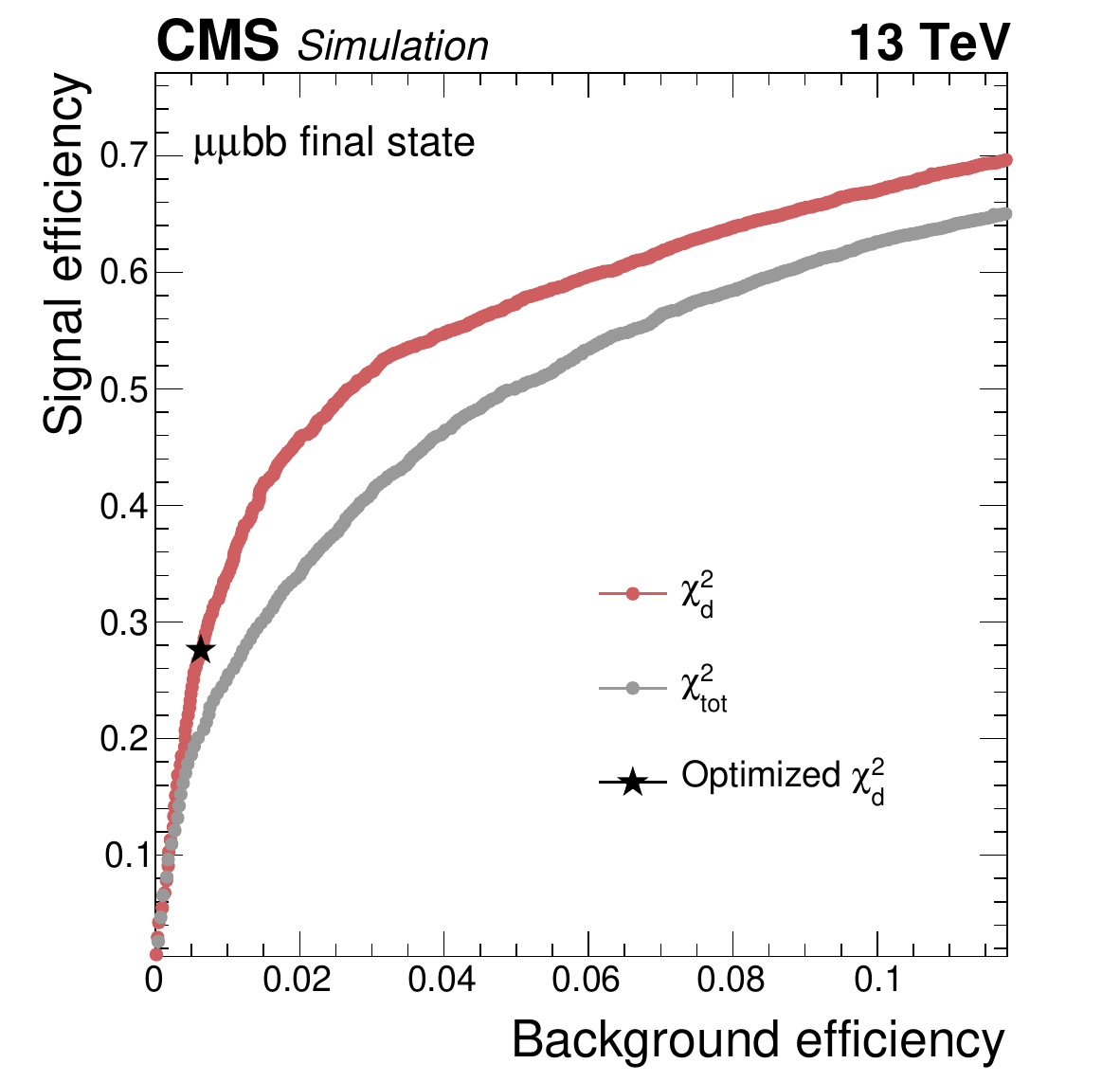}
        \caption{Signal ($\ma=40\GeV$) versus background
          efficiency for different thresholds on $\chit^2$
          (gray) and $\chid^2$ (red) variables. The black star
          indicates signal efficiency versus that of
          background for the optimized $\chid^2$
          requirement. }
        \label{fig:chi2perf}
\end{figure}

The correlation between the \chit components, as well as their bias, depends on \ma. The bias is modeled as
a function of \mmumu and is corrected event by event. After applying this correction, a principal component analysis~\cite{pca_nature} is performed on the bias-corrected variables. The bias-corrected variables are therefore transformed using the eigenvalues, $\lambda_1$ and $\lambda_2$, and eigenvectors $\begin{pmatrix} a\\ b \end{pmatrix}$, of the correlation matrix:
\begin{equation}
\begin{aligned}
\label{eq:corr}
\begin{pmatrix}
\chih\\ \chib
\end{pmatrix}_\mathrm{d} = 
\begin{pmatrix}
\frac{a}{\sqrt{\lambda_1}}&\frac{b}{\sqrt{\lambda_1}}\\ \frac{-b}{\sqrt{\lambda_2}}&\frac{a}{\sqrt{\lambda_2}}
\end{pmatrix}
\begin{pmatrix}
\chih\\ \chib
\end{pmatrix}_\mathrm{c},\\
\chid^2 \equiv \chihd^2 + \chibd^2
\end{aligned}
\end{equation}
with subscripts $\mathrm{d}$ and $\mathrm{c}$, respectively, standing for decorrelated and
bias-corrected components of \chit. The transformation matrix in
Eq.~(\ref{eq:corr}) has three independent parameters,
$a/\sqrt{\lambda_1}$, $a/\sqrt{\lambda_2}$, and $b/a$, that are
modeled as functions of \ma. Figure~\ref{fig:chi2perf} compares the
performance of the selection based on $\chid^2$ and $\chit^2$
variables in terms of the signal ($\ma=40\GeV$) efficiency and
background rejection probability. Based on the optimization studies, events with $\chid^2<1.5$ are selected.

Table~\ref{tab:optYield} summarizes the number of observed events in
data together with the expected yields for the main backgrounds and the signal for
different \ma hypotheses.

\begin{table*}[!ht]
        \topcaption{Event yields in the \mmbb channel for simulated processes and the
          number of observed events in data after applying
          $\chid^2<1.5$. The expected number of simulated events is
          normalized to the integrated luminosity of $\il$. The
          Type III parametrization of 2HDM+S with $\tanb=2$ is used to
          evaluate $\mathcal{B}(\ab\ab\to\mmbb)$.}  
        \centering
        \begin{tabular}{lc}
                \hline Process & Yield \\ \hline {\toppair}+jets &86.3 $\pm$ 2.2
                \\

                DY (10 $< \mellell <$ 50\GeV) & 289.6 $\pm$
                89.5 \\

                DY ($\mellell >$ 50\GeV ) & 200.2 $\pm$ 31.9 \\

                Diboson & 1.5 $\pm$ 0.9 \\

                Single top & 11.4 $\pm$ 1.6 \\ \hline Total expected
                background &589.1 $\pm$ 95.1 \\ Data & 641 \\ \hline
        \end{tabular}
        \\[2ex]
        \begin{tabular}{ccc}
                \hline \multicolumn{3}{c}{Signal for ggH
                  ($\PGm\PGm\PQb\PQb$)} \\
                  $\ma = 20\GeV$& $\ma = 40\GeV$ & $\ma = 60\GeV$ \\ \hline
                  15.4 $\pm$ 0.2 & 18.7 $\pm$ 0.2 & 40.5 $\pm$ 0.3 \\ \hline
        \end{tabular}
        \label{tab:optYield}
\end{table*}

Events are further categorized according to the jet \pt, the {\PQb} tagging score of the jets, and additional
jet activity in the event compatible with the VBF signature. Events containing
at least one of the two selected \PQb jets with $\pt < 20\GeV$ are put in a separate category
(Low\,\pt). This category brings extra sensitivity to the signals with
lower \ma values and contains about 70\% (40\%) background
(${\Pg\Pg}$F signal) events. For the VBF category, events must have at
least two jets, in addition to {\PQb} jets, with $\pt>30\GeV$,
$\abs{\eta}<4.7$, and an invariant mass $\mjj>250\GeV$. About 50\% of
VBF signal events fall in this category. The remaining events are categorized based on the \PQb tagging score of the looser \PQb jet. Three exclusive categories are defined where the second jet passes the loose but fails the medium (TL), passes the medium but fails the tight (TM), and passes the tight (TT) \PQb tagging working point. This categorization relies on the fact that events with genuine and
misidentified \PQb quark jets are distributed differently among those categories. About 20\% of backgrounds as well as the ${\Pg\Pg}$F
signal events fall into the TL category. The TM and TT categories
almost equally receive 20\% of the ${\Pg\Pg}$F signal and 5\% of the
background events. Table~\ref{tab:CatDefinition} summarizes the
categories of the current analysis, whereas the expected yields in
different categories are presented in
Table~\ref{tab:YieldPerCategory}.

\begin{table}[!ht]
        \topcaption{Summary of the categorization requirements in the \mmbb channel. Events in
          these categories contain two muons and two \PQb jets. As stated in the text, L, M, and T
          stand for the loose, medium, and tight \PQb tagging
          criteria, respectively.  }  
        \centering
        \begin{tabular}{lc}
                \hline
                \multicolumn{2}{c}{Categories for selected events}\\
                \hline &\\ Low\,\pt
                & At least one \PQb jet with $\pt<20$\GeV\\ &\\

                \multirow{3}{*}{VBF} & Two additional jets with
                $\pt>30\GeV$,\\ & $\abs{\eta}<4.7$, and
                $\mjj>250\GeV$\\ &\\ TL & Looser \PQb jet
                passes L but fails M \\ TM & Looser \PQb jet passes M
                but fails T \\ TT & Looser \PQb jet passes T
                \\ &\\ \hline
        \end{tabular}
        \label{tab:CatDefinition}
\end{table}

\begin{table*}[!ht]
        \topcaption{The expected yields for backgrounds and different
          signal hypotheses in each category of the \mmbb channel.
          }  
        \centering
        \begin{tabular}{lcccc}
                \hline \multirow{2}{*}{Category}&
                \multicolumn{3}{c}{Signal for ${\Pg\Pg\Hb}$ ($\PGm\PGm\PQb\PQb$)} & \multirow{2}{*}{Expected background}\\
                & $\ma = 20\GeV$& $\ma = 40\GeV$ & $\ma =
                60\GeV$ & \\ \hline

                Low\,\pt & 7.4 $\pm$ 0.1 & 7.3 $\pm$ 0.1 & 17 $\pm$
                0.2 & 421 $\pm$ 88 \\

                VBF & 0.2 $\pm$ ($<$0.1) & 1.0 $\pm$ ($<$0.1) & 1.1 $\pm$ 0.4
                & 5 $\pm$ 2 \\

                TL & 2.1 $\pm$ 0.1 & 2.8 $\pm$ 0.1 & 6.7 $\pm$ 0.1
                & 109 $\pm$ 30 \\

                TM & 2.7 $\pm$ 0.1 & 3.3 $\pm$ 0.1 & 7.7 $\pm$ 0.1
                & 27 $\pm$ 15 \\

                TT & 2.8 $\pm$ 0.1 & 4.2 $\pm$ 0.1 & 8.1 $\pm$ 0.1 &
                28 $\pm$ 11 \\

                \hline Total& 15.4 $\pm$ 0.2 &
                18.7 $\pm$ 0.2 & 40.5 $\pm$ 0.3 & 589 $\pm$ 95 \\
                \hline
        \end{tabular}
        \label{tab:YieldPerCategory}
\end{table*}

In the \ttbb channel, the offline signal event signature constitutes at least one \PQb jet, and depending on the \PGt lepton decay mode, an \emu, an \etau, or a \mutau pair. Any event with an additional electron or muon is rejected to reduce the contribution from DY and multilepton processes. The selection and identification requirements for all objects are discussed in Section~\ref{sec:obj}.

Each final state is subdivided into two categories based on the presence of exactly one \PQb jet or at least two \PQb jets in the event.
Requiring at least two \PQb jets in the event introduces an additional category compared to Ref.~\cite{2016bbtautau}, capable of reconstructing the full signal hypothesis and bringing further signal-to-background discrimination power.
In total there are six event categories, considering the number of \PQb jets and the decay modes of the $\PGt$ leptons.
A deep neural network (DNN) with two hidden layers and 40 nodes is used to discriminate signal from background events in each category.
The DNNs are trained using simulated events.

Kinematic properties of the decay products are utilized to construct variables that are inputs to the DNN training, such as the \pt and transverse mass (\mT) of the leptons and \PQb jets, \pt and $\eta$ of the di-$\PGt$ system, the invariant mass of each system made of a lepton and a \PQb jet, and $\Delta R$ between various combinations of the identified particles.
One of the most important discriminating observables used in the training is the invariant mass of the decay products of the $\PGt$ leptons and the \pt-leading \PQb jet, denoted by $m_{b\PGt\PGt}$.
The $m_{b\PGt\PGt}$ value is typically smaller for signal than for background events.
Similarly, angular separation and other invariant mass variables can be reconstructed with different combinations of the four final-state particles, employing the correlation between the resonance decay products.
The \mT between an \Pe or \PGm and \ptmiss is one of the discriminating variables and is defined as
\begin{equation}
        \mT (\Pe/\PGm, \ptmiss) \equiv \sqrt{2 \ptem \ptmiss \left[1 - \cos(\Delta\phi)\right]},
\end{equation}
where \ptem is the transverse momentum of the lepton and $\Delta\phi$ is the azimuthal angle between the lepton direction and \ptvecmiss.
Events from {\toppair}+jets and misidentified \tauh backgrounds, such as \wjets, have larger \ptmiss, thus result in higher \mT values.

Another variable useful in the training is \dzeta, defined as
\begin{equation}
        \dzeta \equiv \pzeta - 0.85 \pzeta^{\text{vis}},
\end{equation}
where the bisector of the directions of the visible $\PGt$ decay products transverse to the beam direction is denoted as the $\zeta$ axis.
The quantity \pzeta is defined as the component of the \ptvecmiss along the $\zeta$ axis, and $\pzeta^{\text{vis}}$ to be the sum of the components of the lepton transverse momentum along the same direction~\cite{Abulencia:2005kq}.
The $\PZ\to\PGt\PGt$ background corresponds to large \dzeta values because the \ptmiss is approximately collinear to the $\PGt\PGt$ system.
The {\toppair}+jets events tend to have small \dzeta values due to a large \ptmiss that is not aligned with the $\PGt\PGt$ system.
The signal has intermediate \dzeta values because the \ptmiss is approximately aligned with the $\PGt\PGt$ system, but its magnitude is small.

For events in the category with two or more \PQb jets, a variable can be constructed to measure the difference between the invariant mass of the two \PQb jets and the invariant mass of the $\PGt\PGt$ system ($m_{\PGt\PGt}$):
\begin{equation}
        \Delta \ma \equiv (m_{\PQb\PQb}-m_{\PGt\PGt})/m_{\PGt\PGt}.
\end{equation}
This variable is of particular interest since it peaks at zero for signal events. 
The $m_{\PGt\PGt}$ distribution reconstructed with the SVfit algorithm~\cite{svfit} is used to test the presence of signal, and thus is not directly included as an input to the DNN.

\begin{figure}[!htb]
  \centering
  \includegraphics[width=0.42\textwidth]{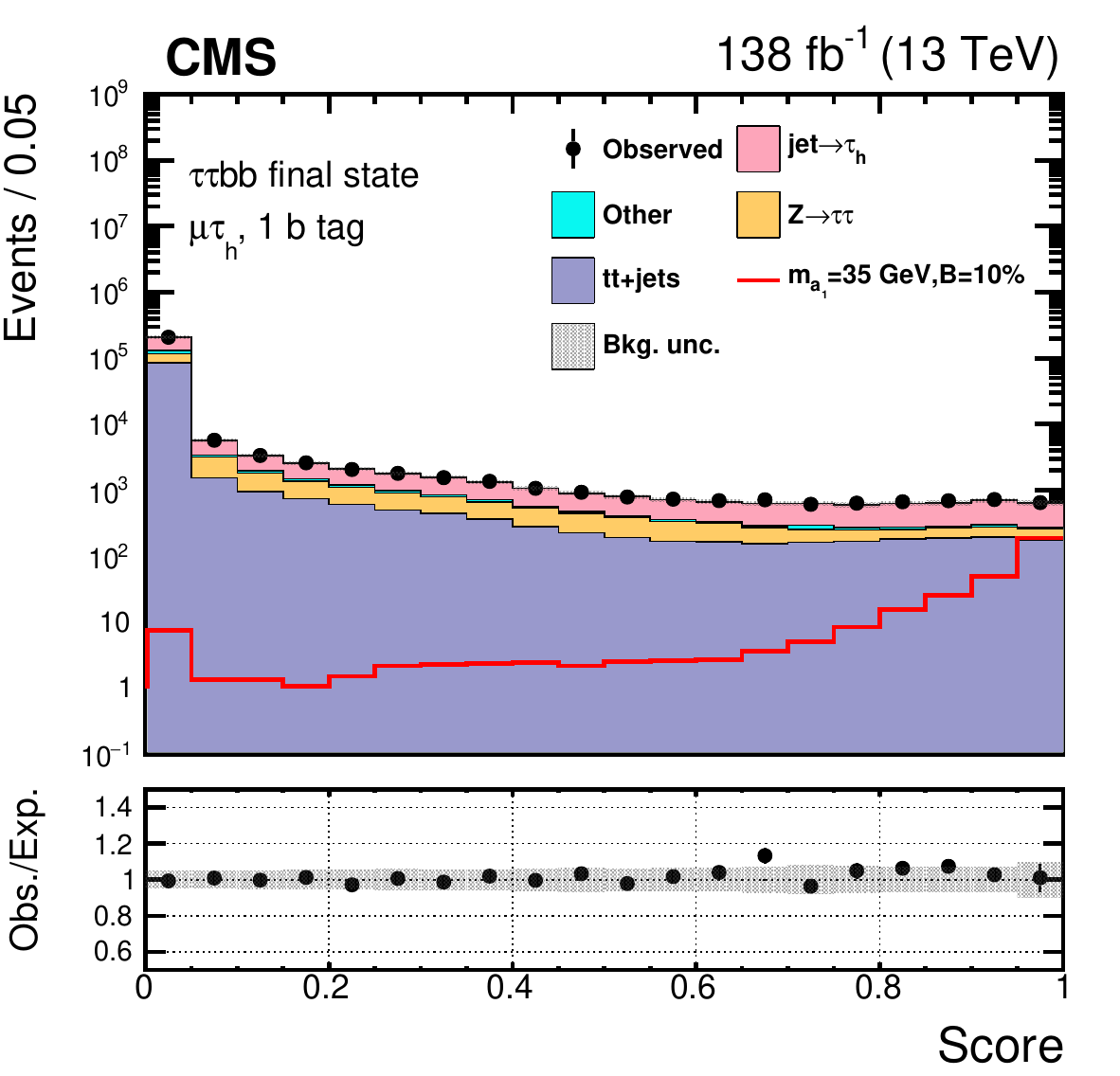}
  \includegraphics[width=0.42\textwidth]{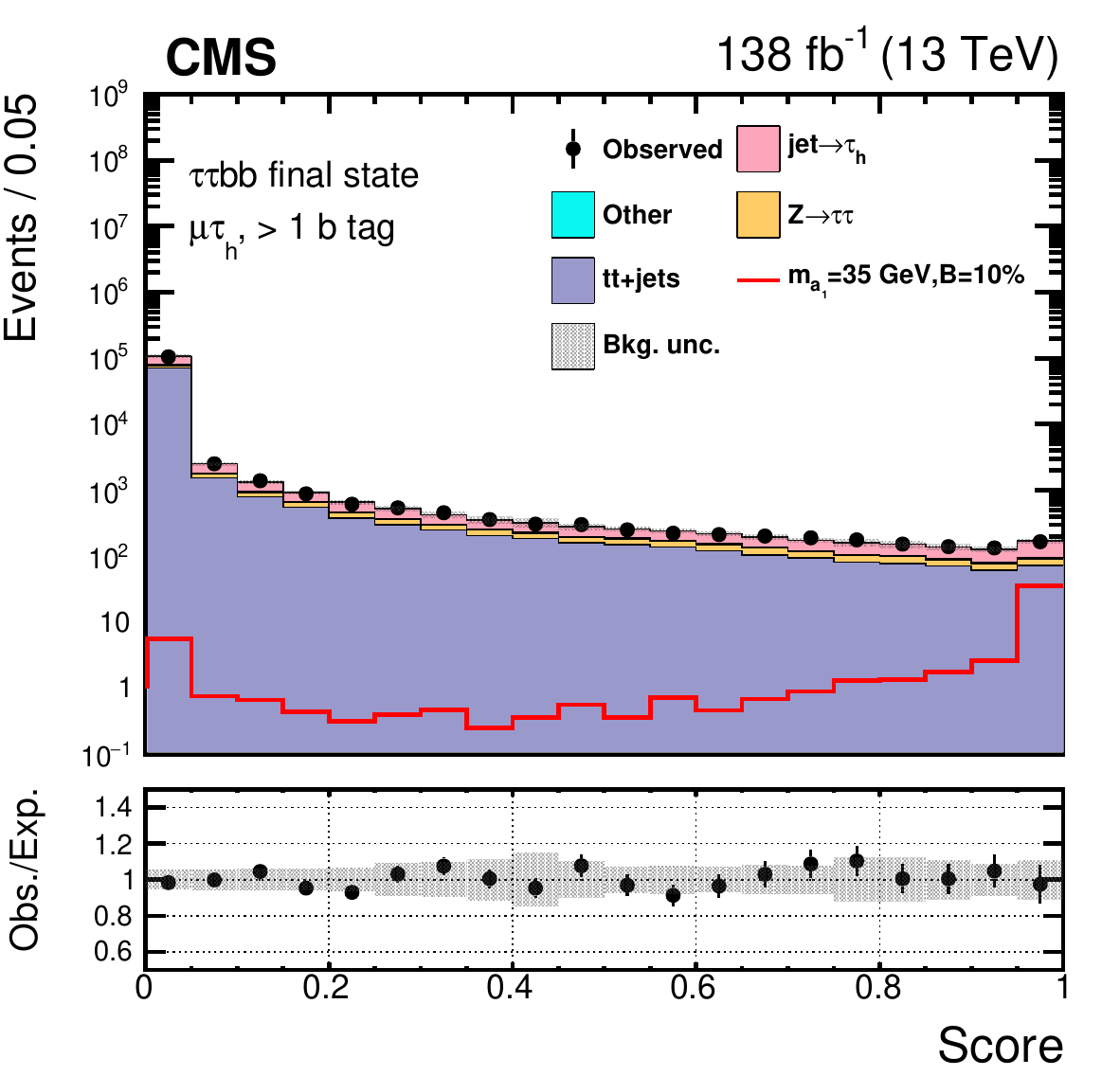}
  \caption{
  Pre-fit distributions of the DNN score for the \mutau channel divided into events with one (\cmsLeft) or at least two (\cmsRight) \PQb jets.
  The shape of the $\Hb\to\ab\ab$ signal, where $\ma = 35\GeV$, is indicated assuming $\mathcal{B}(\Hb\to\ab\ab\to\ttbb)$ to be 10\%.
  The lower panel shows the ratio of the observed data to the expected yields.
  The gray band represents the unconstrained statistical and systematic uncertainties.
  \label{fig:postfit_dnn}}
\end{figure}

Figure~\ref{fig:postfit_dnn} shows, as an example, the DNN score distributions in the \mutau channel separated for events with one or at least two \PQb jets.
The distributions are obtained by comparing the estimated signal and background distributions of the DNN score to that of the data before the fit described in Section~\ref{fitmethod} (pre-fit).

In each category, subregions are defined using a threshold on the DNN score.
The expected limits are scanned by varying the DNN thresholds to obtain the highest sensitivity to the simulated signal.
This optimization method also ensures that the expected number of background events in each subregion is large enough to perform the final likelihood fit of the $m_{\PGt\PGt}$ distribution.
There are three SRs for events containing one \PQb jet: SR1, SR2, and SR3, whereas events with two \PQb jets are divided into two categories: SR1 and SR2. The only exception is the \etau final state in the two-\PQb-jet category where no significant gain was observed when adding a second signal region.
The remaining subregion containing events with the lowest DNN scores is used as a control region (CR) to constrain various background normalizations in the final likelihood fit.
\section{Background estimation}
The presence of a \mmbb signal is expected to appear as a peak over the \mmumu distribution centered at \ma. The background shape and its normalization in this channel are collectively determined from data with no reference to simulation. Different parameterizations of polynomials are used to model the \mmumu distribution in data of every category, separately. For each group of models, a maximum degree of the polynomial, determined through statistical tests, is imposed. This is to ensure that the data are not overfit. Parameters of every selected model vary within their uncertainties in the final fit to extract the signal strength, defined as the ratio of the observed signal rate to that predicted by the SM. The latter uses the discrete profiling method~\cite{Dauncey:2014xga,Khachatryan:2014ira,Aad:2015zhl} where every functional form of the selected background models is treated as a discrete nuisance parameter. Along with the determination of the signal strength, one of the background models, its parameters, and the corresponding normalization are determined by the fit, as described in Section~\ref{fitmethod}.

A major background contribution to the \ttbb channel is $\PZ\to\PGt\PGt$, which is estimated from data using an embedding technique~\cite{embedding}. The method is based on the reconstruction of $\PZ\to\PGm\PGm$ events in data where the muons are replaced with simulated \PGt leptons with the same kinematic properties. In comparison with the simulation of the $\PZ\to\PGt\PGt$ process, this technique allows a more accurate description of variables related to \ptmiss and jet activity. The embedded sample also estimates other SM processes with two genuine \PGt leptons, such as {\toppair}+jets and Diboson.

The QCD multijet contribution to the \emu\ final state of the \ttbb channel is estimated using the data in a sideband (SB) region with same-sign \emu pairs. The event selection in the SB region is otherwise identical to that in the \emu SRs. The contributions of other processes in the SB are taken from simulation and subtracted from the data. The resulting number of data events in the SB is scaled by the ratio of the expected multijet contribution in the SR to the expected multijet contribution in the SB. Scale factors are calculated in data orthogonal to the SR, as functions of the jet multiplicity and the $\Delta R$ separation between the electron and the muon, in order to account for possible kinematic differences between the two regions.

Backgrounds with hadronic jets that are misidentified as \tauh candidates contribute significantly to \etau and \mutau final states and are estimated from data. This background includes the \wjets, QCD multijets, and {\toppair}+jets processes with at least one top quark decaying to hadrons. In a data sideband region, events are required to pass all the baseline \etau/\mutau selection criteria, but fail the \tauh\ isolation. The data in this SB are reweighted with a factor $f/(1-f)$, where $f$ is the probability for a jet to be misidentified as a \tauh candidate and is evaluated as a function of the $\pt(\tauh)$. The $\PZ\to\PGm\PGm$+jets events in data are used to measure the misidentification probability. The final state must contain a dimuon pair compatible with the decay of the \PZ boson, as well as a \tauh candidate. Simulation is used to subtract from data the contribution from events with a genuine \tauh lepton. The measurement is done separately for the \etau\ and \mutau final states. This is because the antilepton discrimination working points in the \tauh identification change depending on the lepton selected is an electron or a muon~\cite{CMS:2022prd}. The difference between the two fake rate measurements is observed to be around 10\%. The misidentification probability also depends on the jet multiplicity, which characterizes the hadronic activity in the event.

Another dominant background is {\toppair}+jets, which has to be carefully estimated from simulation.
Because {\toppair}+jets events with two genuine \PGt leptons in the final state are an irreducible contribution to the embedded sample described above, the {\toppair}+jets background estimate from simulation described here does not include these events.
It also does not include {\toppair}+jets events in which a reconstructed \tauh candidate arises from a simulated jet, as the estimation of the misidentified \tauh background is derived from data SBs, as described above.
The normalization of backgrounds is free to vary within a range limited by the a priori uncertainty estimates in the final fit for the signal extraction.

The presence of a \ttbb signal is expected to appear as a peak over the \mtt distribution centered at \ma. A fit to the \mtt distribution is performed simultaneously in the SRs and CRs described in Section~\ref{sec:evt}.
\section{Signal extraction}
\label{fitmethod}
In the \mmbb final state, an unbinned maximum likelhood fit to the data \mmumu distributions is carried out simultaneously in all event categories.
The fit is performed in the range $15<\mmumu<62.5\GeV$, using parametric models for signal and background.
The parametric model of the signal is a weighted sum of a Voigt profile and a Crystal Ball (CB) function~\cite{CrystalBallRef}, where the mean values of the two are constrained to be identical~\cite{2019398}.

Simulated samples are used to determine the parameters of the signal model that may depend on \ma. The studies are performed separately on signal samples simulated for different years. This is to account for the effect of muon reconstruction details on the signal model in different data-taking periods. Most of the parameters are found to be independent of \ma and fixed in the final fit. Only the resolutions of the Voigt profile and CB function demonstrate linear variation with the pseudoscalar mass. The slope of the linear models are floating parameters in the signal extraction fit. In each category, contributions from different years are normalized considering the signal selection efficiency and acceptance, and are used to construct the expected signal distribution in data. The expected signal efficiency and acceptance are interpolated for \ma values not covered by simulation.

To evaluate the background contribution, every selected functional form is treated as a discrete nuisance parameter as discussed earlier. In addition, the parameters of every model, as well as the normalization, are part of the background parameter space. A likelihood $\mathcal{L}$ is constructed using the signal and the background models in all categories, including systematic uncertainties associated with the signal, as nuisance parameters. In the minimization process of the negative logarithm of the likelihood, the discrete profiling method chooses a best fit background model as the physics parameter of interest, the signal strength, varies. The method incorporates the systematic uncertainty in the background model by taking the envelope of the models provided to the fit.

\begin{figure*}[h!tp]
  \centering
  \includegraphics[width=0.42\textwidth]{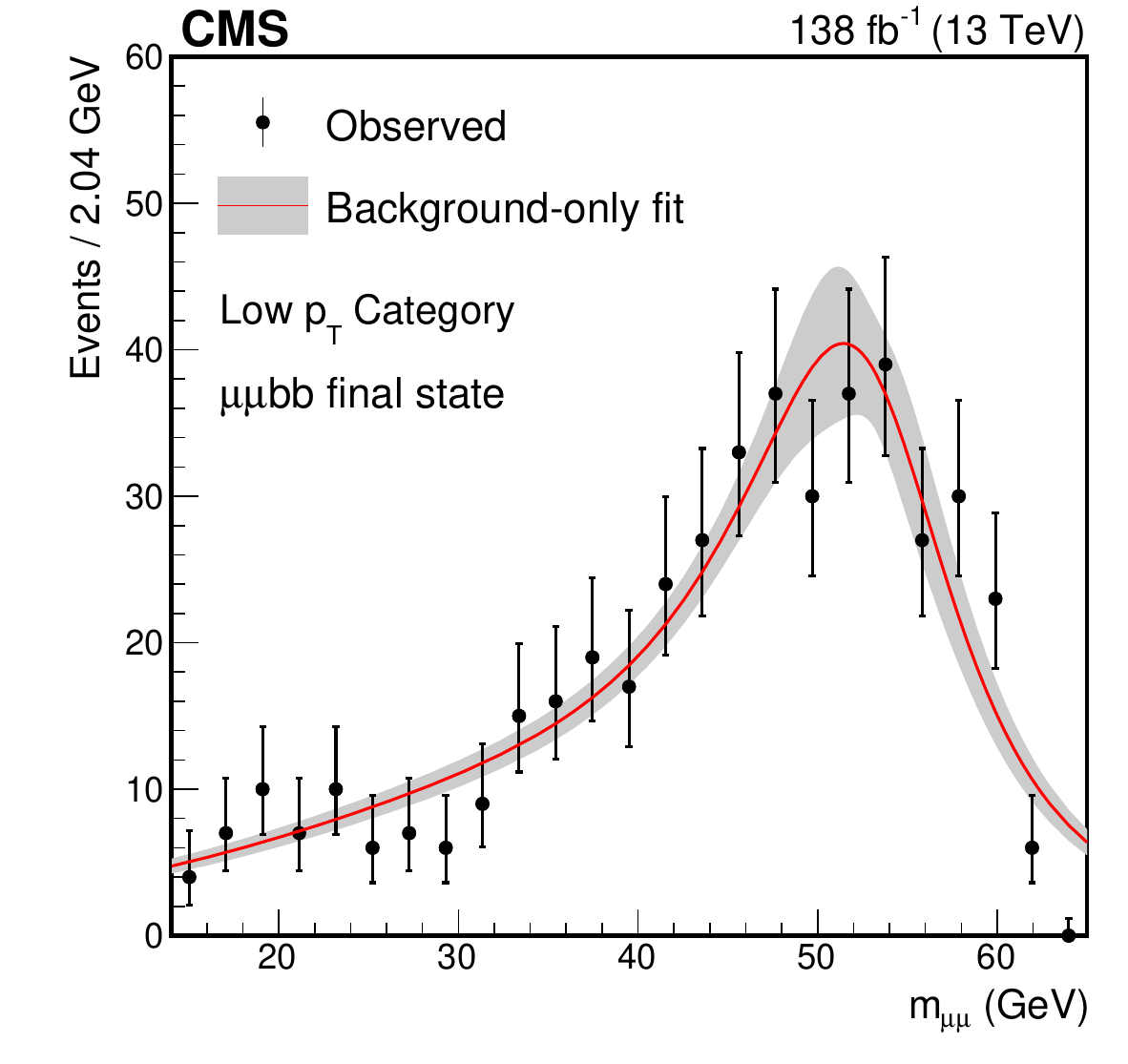}
  \includegraphics[width=0.42\textwidth]{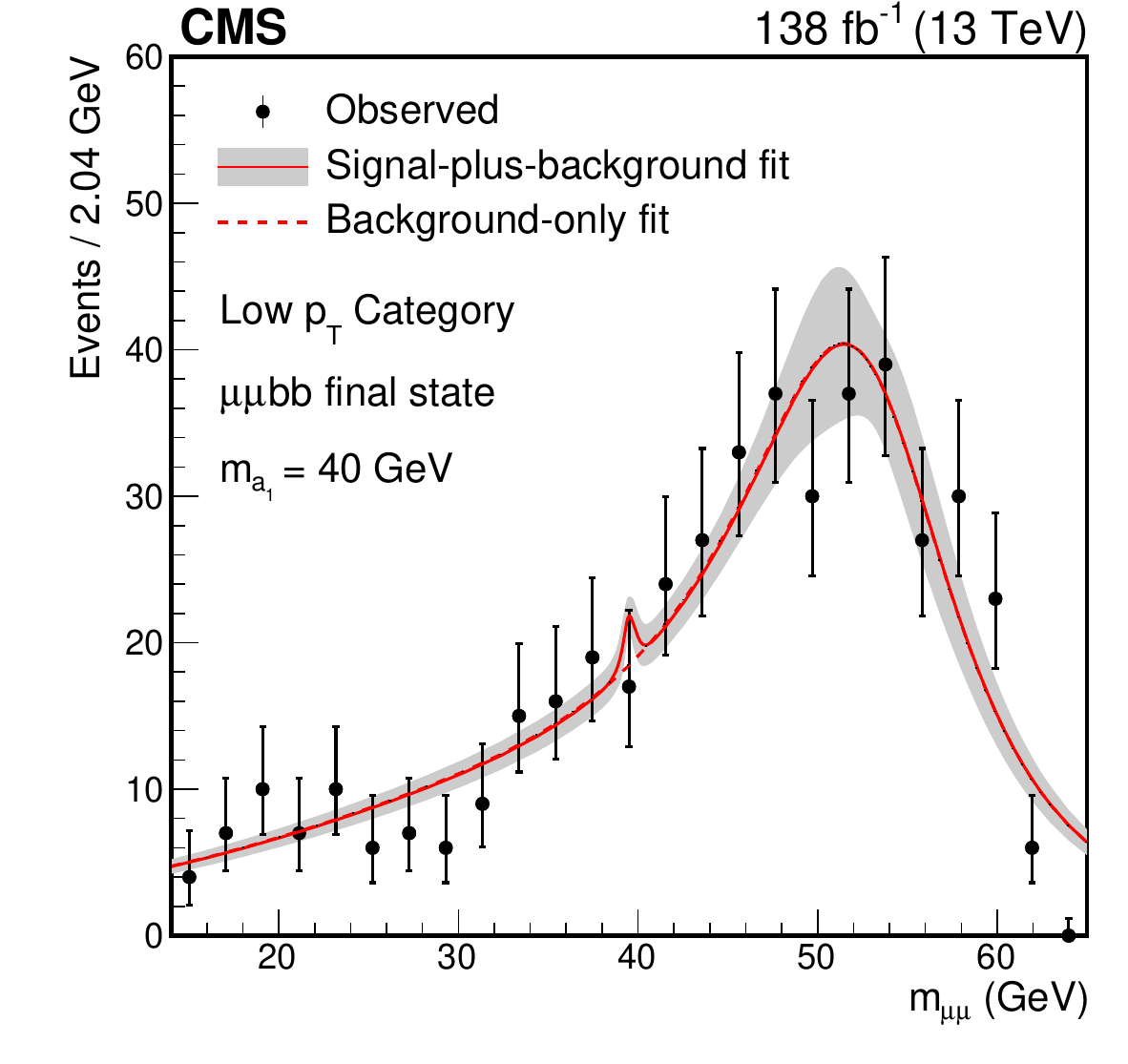} \\
  \includegraphics[width=0.42\textwidth]{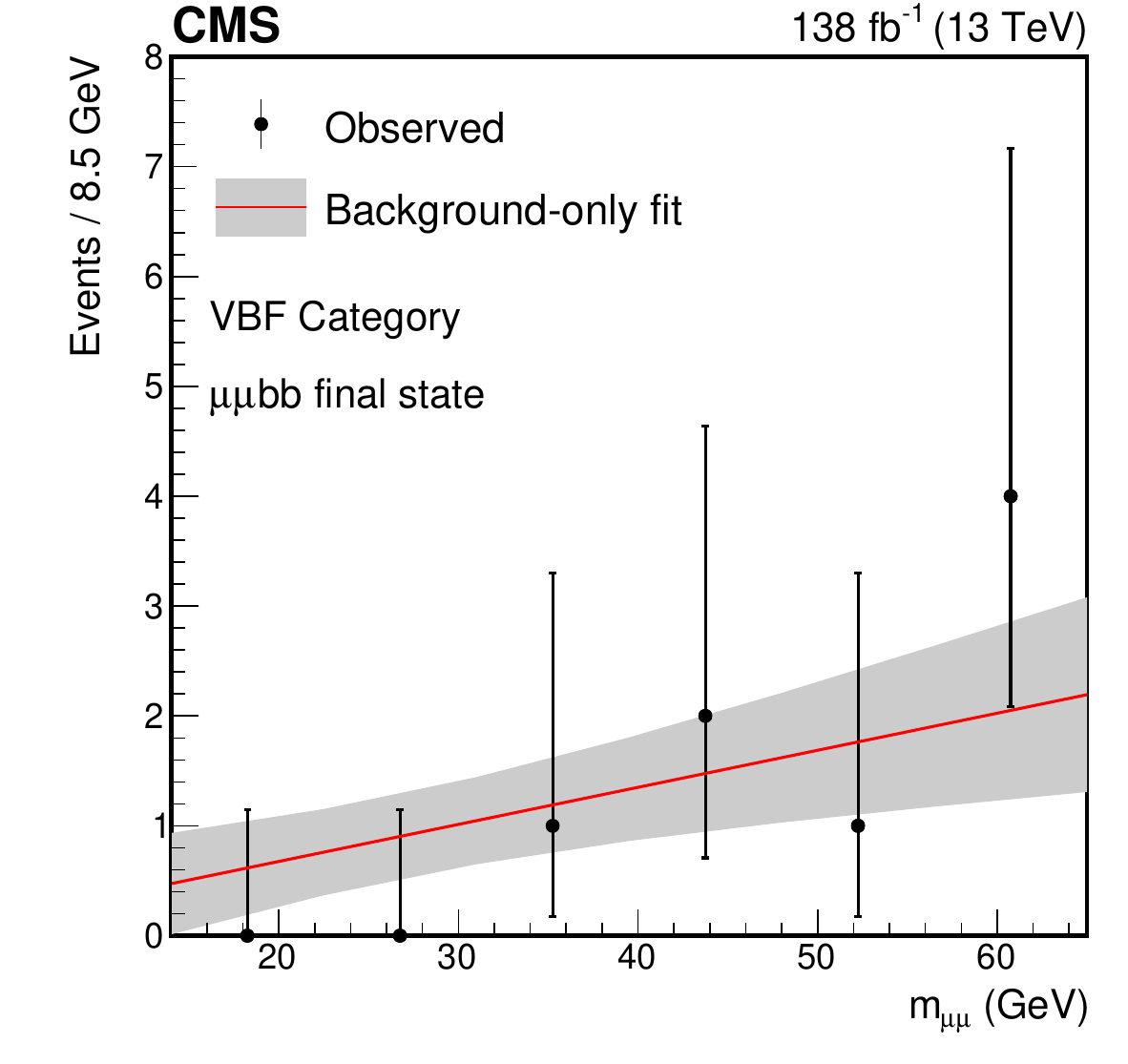}
  \includegraphics[width=0.42\textwidth]{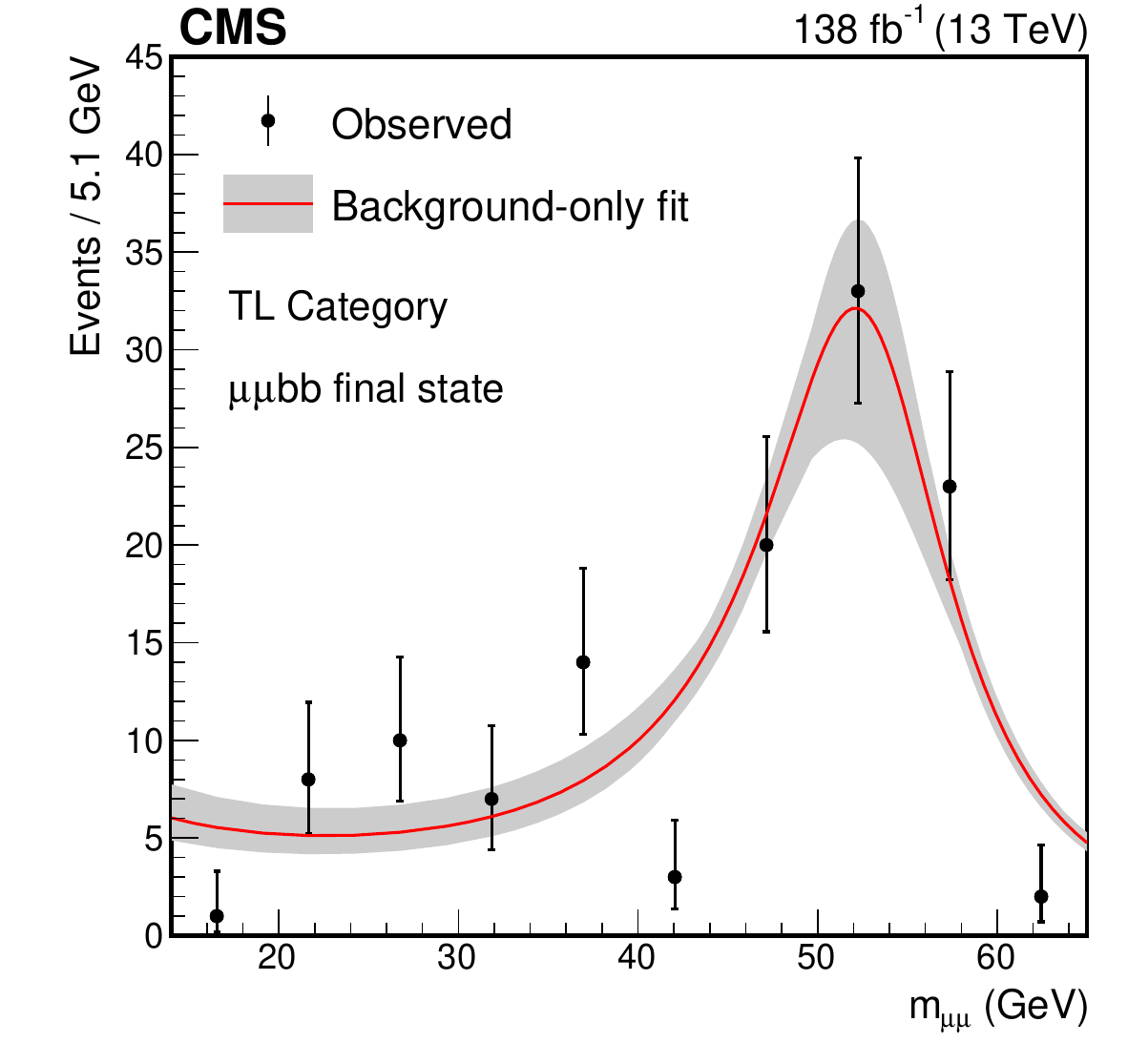} \\
  \includegraphics[width=0.42\textwidth]{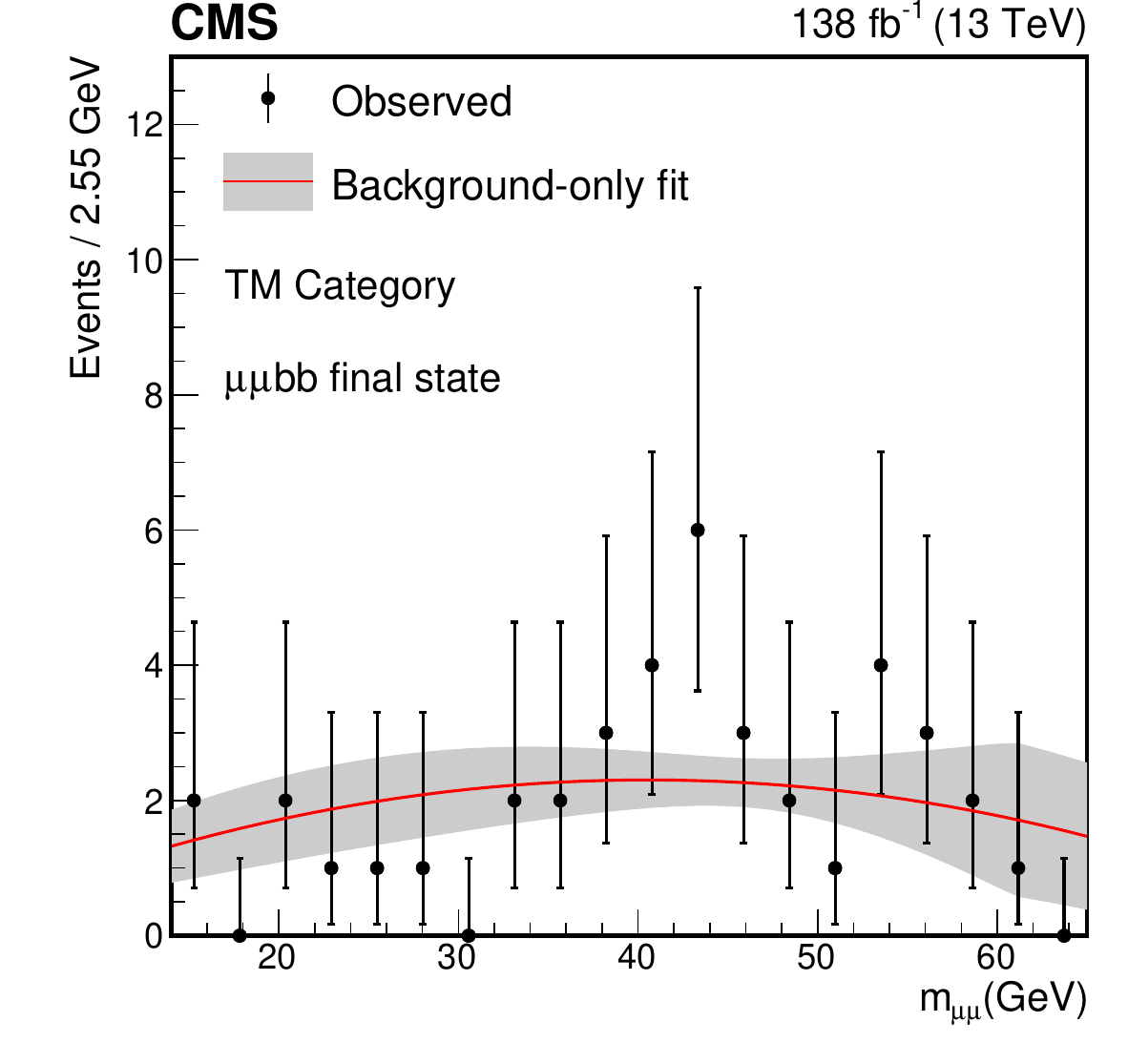}
  \includegraphics[width=0.42\textwidth]{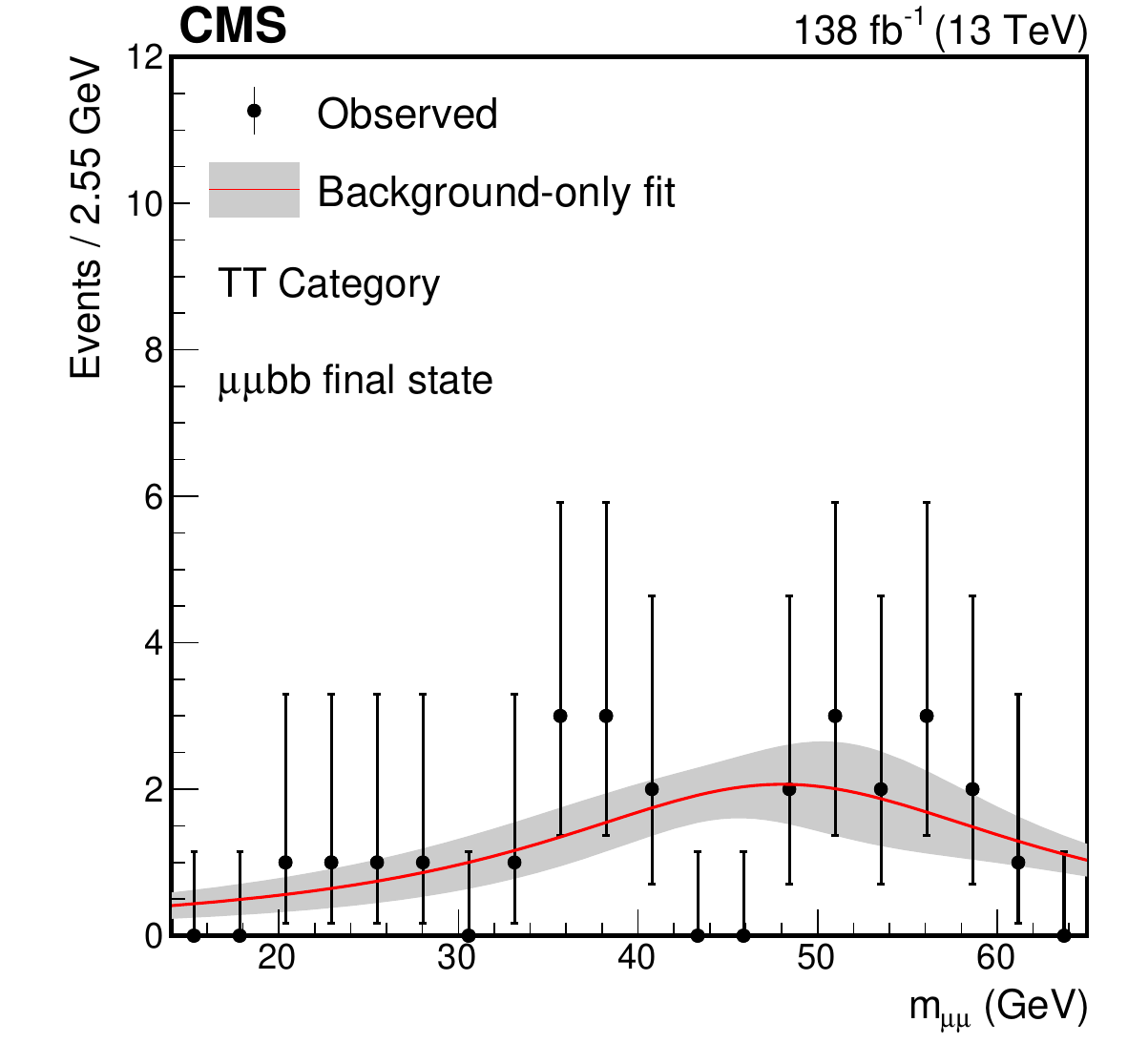}
  \caption{The best fit background models for the \mmbb channel together with a 68\% CL uncertainty band from the fit to the data under the background-only hypothesis for the (upper left) Low\,\pt category, (middle left) VBF category, (middle right) TL category, (lower left) TM category, and (lower right) TT category. For comparison, the signal-plus-background is shown for the (upper right) Low\,\pt category for a signal with $\ma = 40\GeV$. The expected signal yield is evaluated assuming the SM production of the Higgs boson and $\mathcal{B}(\ab\ab\to\mmbb)=0.2\%$, as predicted in the Type III 2HDM+S with $\tanb=2$. The bin widths depend on statistics, irrelevant for the final fit.}
    \label{fig:postfitmmbb}
\end{figure*}

In practice, a penalty term is added to the likelihood to account for the number of free parameters in the final background model. The penalized likelihood, $\widetilde{\mathcal{L}}$, is a function of the measured signal strength, $\mu$, the continuous nuisance parameters, $\vec{\theta}$, and the background models, $\vec{b}$. The penalized likelihood ratio is defined as
\begin{equation}
        -2 \ln \frac{\widetilde{\mathcal{L}}({\text{data}}|\mu, \hat\theta_\mu, \hat b_\mu)}{\widetilde{\mathcal{L}}({\text{data}}|\hat{\mu}, \hat{\theta}, \hat{b})},
\end{equation}
with the numerator being the maximum $\widetilde{\mathcal{L}}$ for a given $\mu$ at the best fit values of nuisance parameters and background functions.
The denominator is the global maximum of $\widetilde{\mathcal{L}}$, obtained at $\mu=\hat{\mu}$, $\theta = \hat{\theta}$, and $b = \hat{b}$. The background function maximizing $\widetilde{\mathcal{L}}$ at any $\mu$ is used to derive the confidence interval on $\mu$ at any $m_{\mathrm{a_{1}}}$~\cite{Dauncey:2014xga}. It is verified that the fit is unbiased using studies where signals at several \ma values are injected with different strengths. The relative change in signal strength is found to be less than $10^{-4}$. The best fit background models together with their uncertainties are shown in Fig.~\ref{fig:postfitmmbb} for all event categories in the \mmbb analysis.

\begin{figure}[!htbp]
  \centering
  \includegraphics[width=0.42\textwidth]{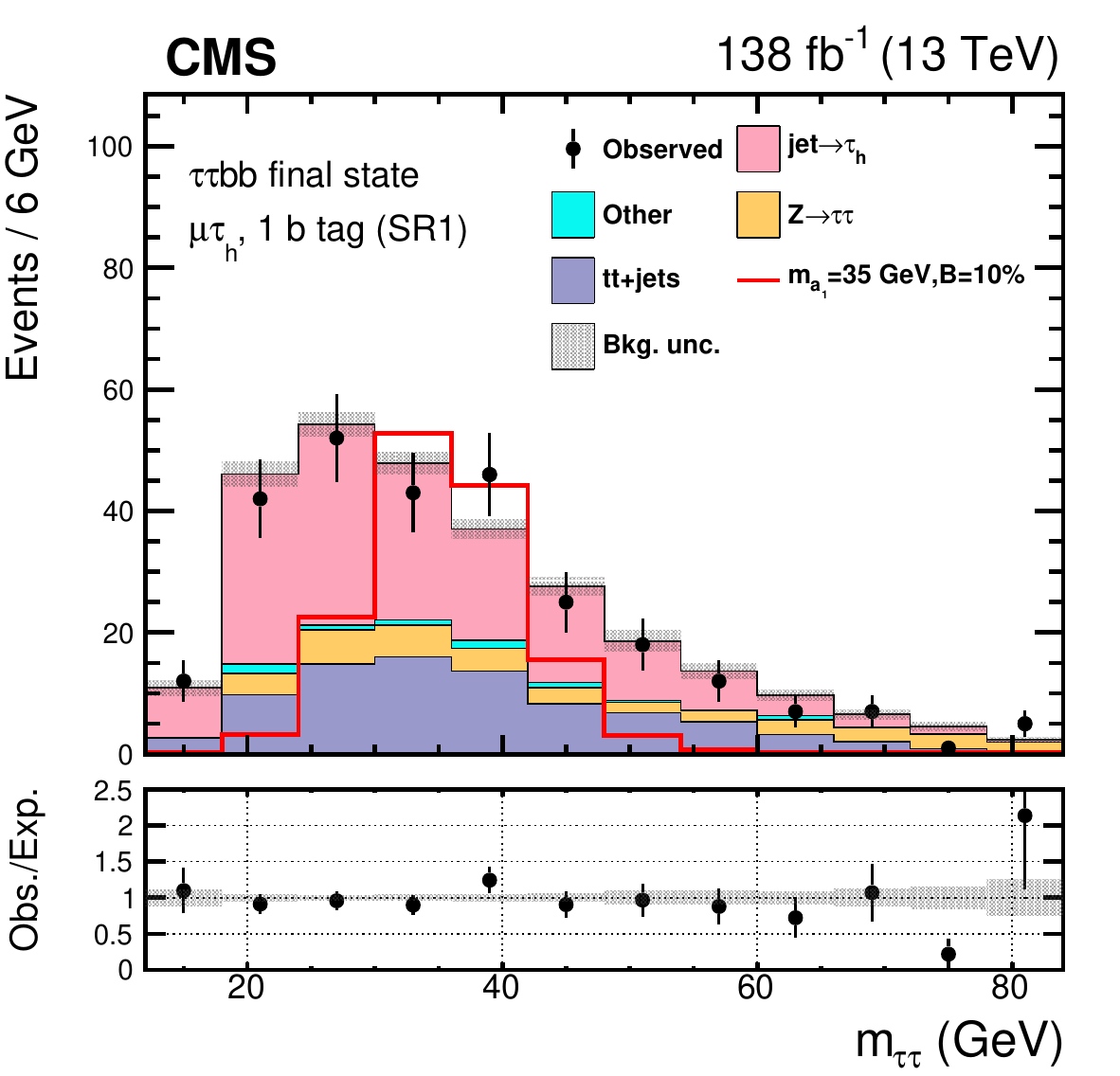}
  \includegraphics[width=0.42\textwidth]{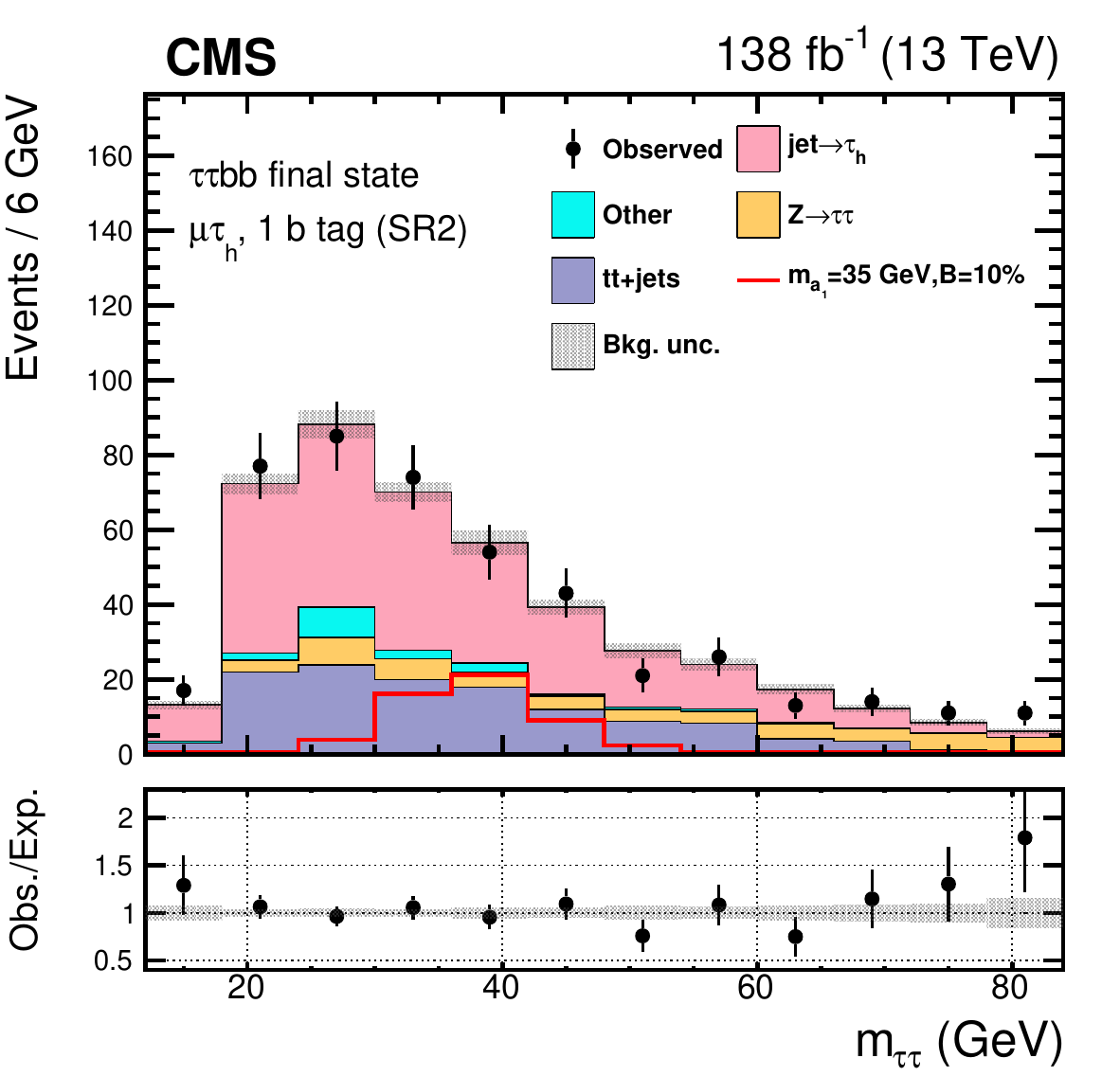} \\
  \includegraphics[width=0.42\textwidth]{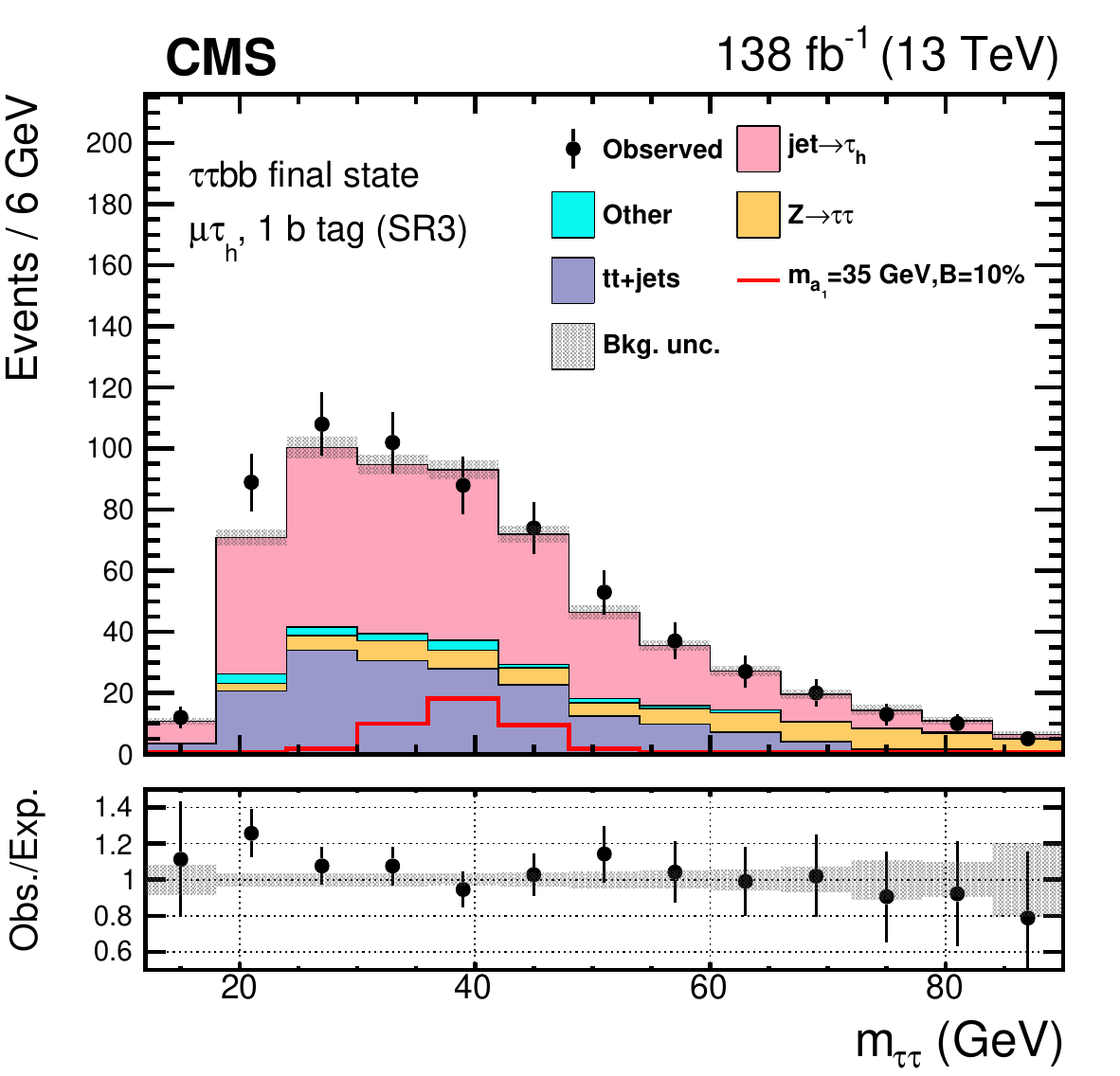}
  \caption{Post-fit distributions of \mtt for the \mutau channel signal regions in events with exactly one \PQb tagged jet: SR1 (\cmsULeft), SR2 (\cmsURight), and SR3 (lower).
  The shape of the $\Hb\to\ab\ab$ signal, where $\ma = 35\GeV$, is indicated assuming $\mathcal{B}(\Hb\to\ab\ab\to\ttbb)$ to be 10\%.
  \label{fig:postfit_mtt_mutau_1b}}
\end{figure}

\begin{figure}[!htb]
  \centering
  \includegraphics[width=0.42\textwidth]{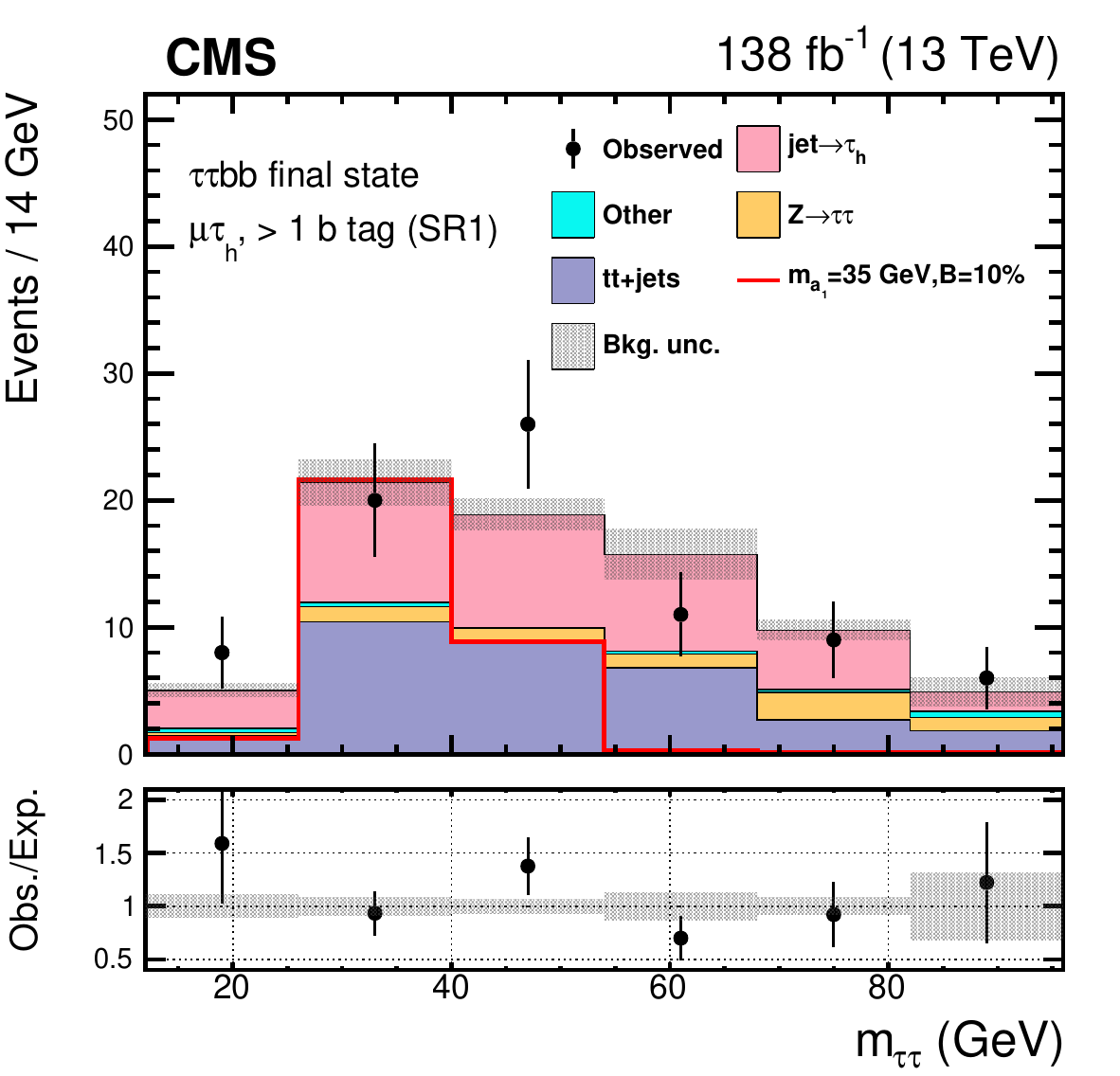}
  \includegraphics[width=0.42\textwidth]{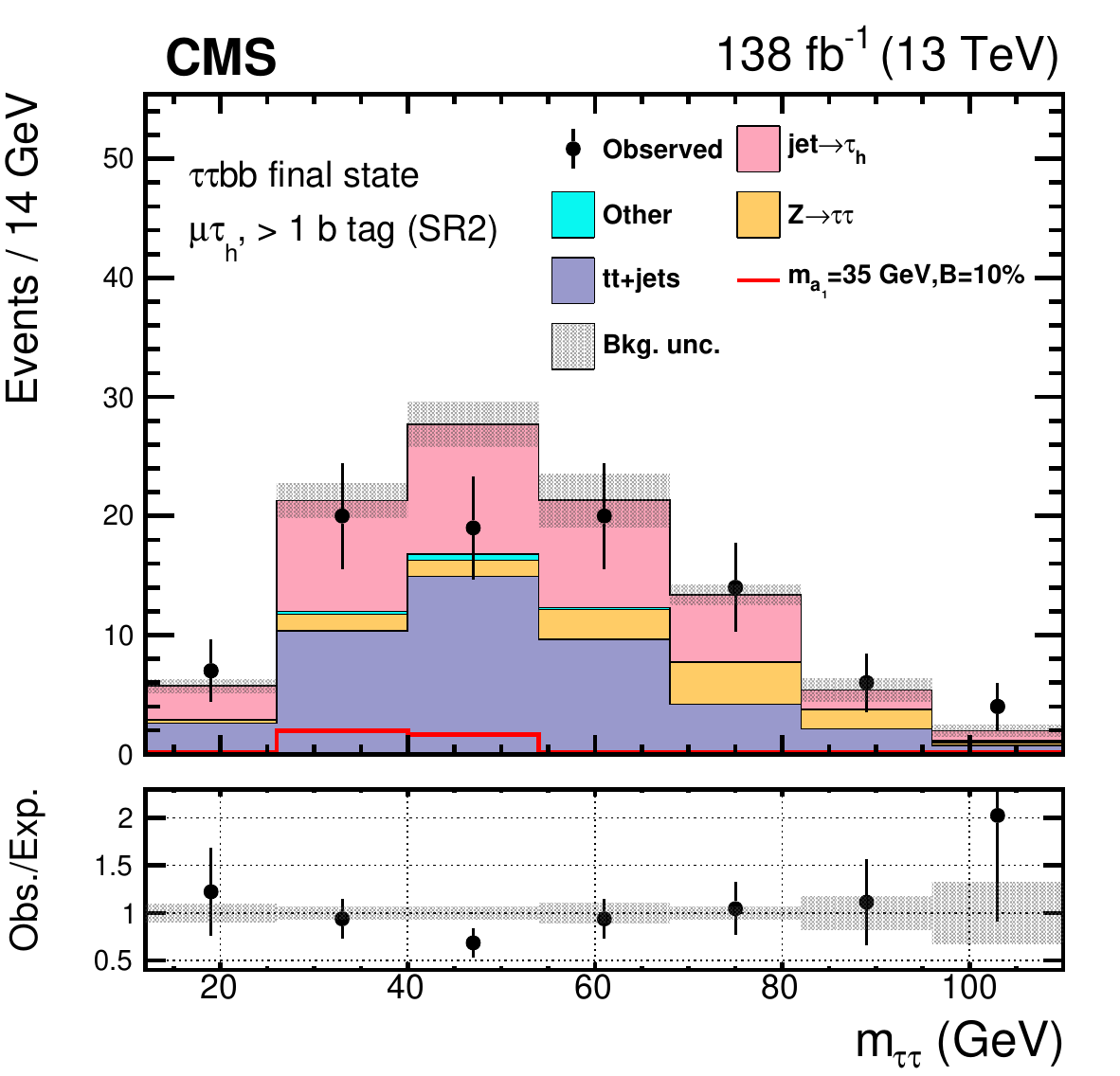}
  \caption{Post-fit distributions of the \mtt for the \mutau channel signal regions in events with at least two \PQb tagged jets: SR1 (\cmsLeft) and SR2 (\cmsRight).
  The shape of the $\Hb\to\ab\ab$ signal, where $\ma = 35\GeV$, is indicated assuming $\mathcal{B}(\Hb\to\ab\ab\to\ttbb)$ to be 10\%.
  \label{fig:postfit_mtt_mutau_2b}}
\end{figure}

\begin{figure}[!htb]
  \centering
  \includegraphics[width=0.42\textwidth]{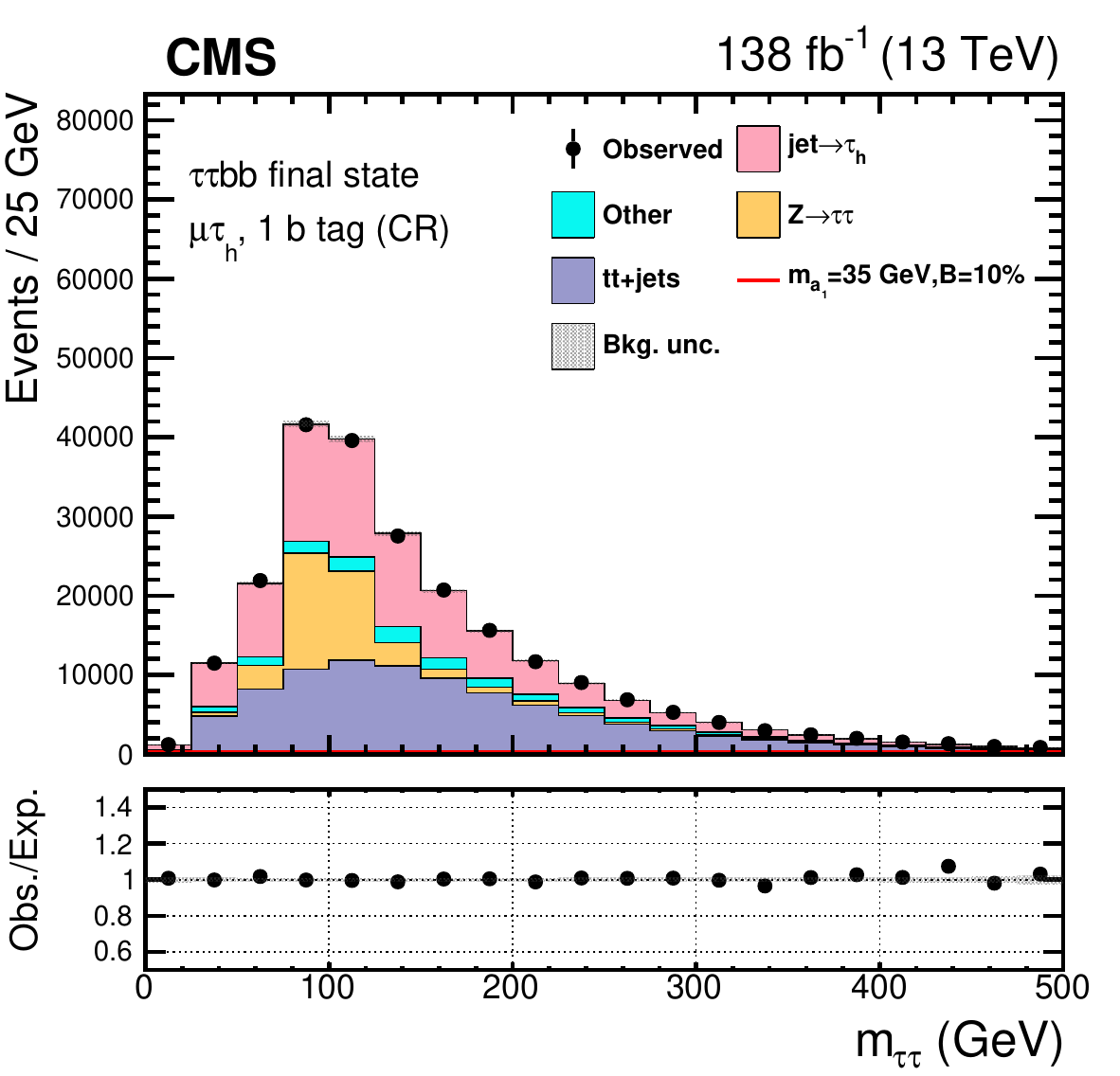}
  \includegraphics[width=0.42\textwidth]{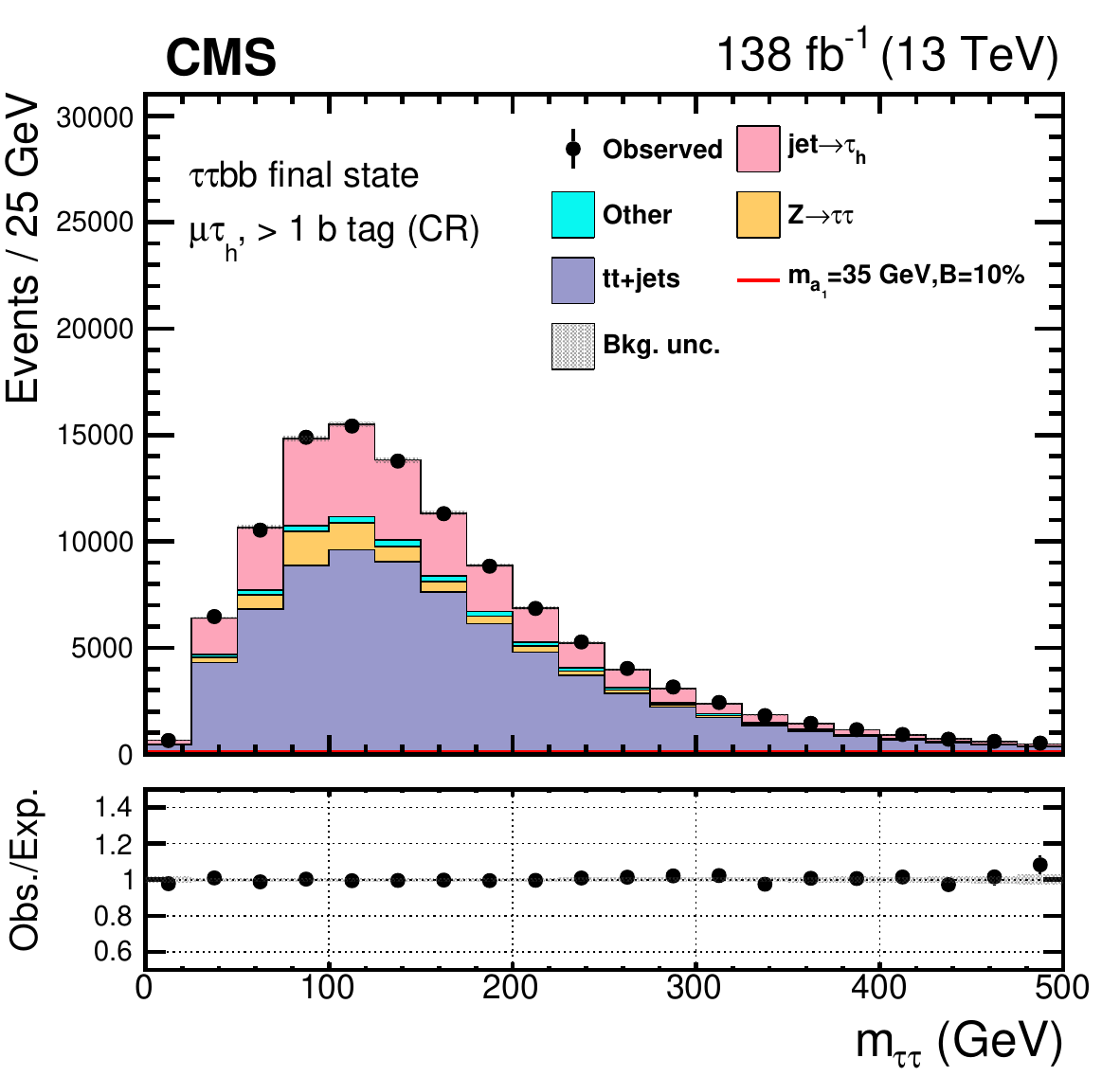}
  \caption{Post-fit distributions of the \mtt for the \mutau channel control regions in events with exactly one \PQb tagged jet (\cmsLeft) and at least two \PQb tagged jets (\cmsRight).
  The contamination from the $\Hb\to\ab\ab$ signal, where $\ma = 35\GeV$, is barely visible assuming $\mathcal{B}(\Hb\to\ab\ab\to\ttbb)$ to be 10\%.
  \label{fig:postfit_mtt_mutau_cr}}
\end{figure}

In the $\ttbb$ channel, a binned maximum likelihood fit is performed on the \mtt distribution with systematic uncertainties included as nuisance parameters.
The subregions of event categories from all final states are included in a simultaneous fit. Figures~\ref{fig:postfit_mtt_mutau_1b}--\ref{fig:postfit_mtt_mutau_cr} show the post-fit \mtt distributions in different subregions and categories for the \mutau final state.


The limits and confidence intervals are obtained using the modified frequentist CL$_{\mathrm{s}}$ approach~\cite{Junk1999435,0954-3899-28-10-313} with an asymptotic approximation to the distribution of the profile likelihood ratio test statistic~\cite{Cowan2011}. Pseudoscalar masses between 12 and 60\GeV are considered using simulated samples described in Section~\ref{sim}.

The \mmumu and \mtt expected distributions are compared to data in a combined fit, integrating over the \ab decay modes. Integrating over \ab decays makes the combination model dependent since the branching fraction of \ab to fermion pairs depends on the model. The 2HDM+S and the theoretical predictions of Ref.~\cite{Haisch:2018kqx} are used for the branching fractions of \ab to muons, \PGt leptons, and \PQb quarks which are fixed in the fit. The selected events are mutually exclusive in the two analyses as events with an extra muon and/or electron are vetoed in the \ttbb selection. A correlation model is employed between the two analyses for the systematic uncertainties that are in common.
\section{Systematic uncertainties}
\label{systematics}
The sensitivity of the two analyses, \mmbb and \ttbb, is mainly affected by the uncertainties arising from the finite size of the data sample. Nevertheless, several sources of systematic uncertainties are included in the determination of the results. Most of the systematic uncertainties are common between the two analyses, although their impact on the result may differ. In this class of uncertainties fall those associated with the modeling and acceptance of the signal, including the PDFs, the strong coupling constant, and the renormalization and factorization scales. In addition, experimental uncertainties associated with, \eg, the jet energy calibrations, \PQb tagging, and muon reconstruction and identification are in common between the two analyses, although the uncertainties related to the background estimations are not. In the \mmbb analysis, uncertainties associated with the parameters of the dimuon resonance model in the signal are taken into account.

The unbinned maximum likelihood fit of the \mmbb analysis accounts for the shape uncertainties in a different way. The impact of systematic variations is found to be negligible on the parametric model of the signal for all \ma hypotheses. On the other hand, the modeling of the \mmumu resolution with \ma (discussed in Section~\ref{fitmethod}) has an uncertainty that is included in the fit with a Gaussian profile. Uncertainties associated with the background model are evaluated by means of the discrete profiling method as described earlier and contribute to the statistical uncertainty of the result. Depending on the signal mass hypothesis, they constitute about 10--25\% of the total uncertainty in the \mmbb results. Contributions from uncertainties in the signal efficiency and acceptance are significantly smaller. In the following, details are provided for several sources of uncertainties.

All uncertainties are included as nuisance parameters in the final fit for the signal extraction. Uncertainties affecting the event yields in categories, \ie, normalization uncertainties, are assigned via multiplicative corrections, with a log-normal probability density function. In the binned maximum likelihood fit of the \ttbb analysis, nuisance parameters that modify the shapes of the \mtt distributions are assumed to have a Gaussian profile.
This means that for every nuisance parameter of this type, two alternate distributions are provided to the fit: one with the distribution resulting from an increase of the nuisance parameter by one standard deviation and the other with the distribution resulting from a decrease by one standard deviation.
The dominant systematic uncertainty is found to be associated with the signal model, followed by the normalization of the QCD multijet background in the \emu final state and the uncertainties in the {\toppair}+jets cross section.

\textbf{Integrated luminosity}: the integrated luminosity of the data recorded by CMS for physics analyses is evaluated separately for different years of the Run~2 data taking~\cite{CMS-LUM-17-003,CMS-PAS-LUM-17-004,CMS-PAS-LUM-18-002}. The uncertainty in the measured integrated luminosity of a given year has a component that is uncorrelated across the years. It amounts to 1.0, 2.0, and 1.5\%, for the 2016, 2017, and 2018 periods, respectively. Another component is correlated across all three years and is 0.6\% in 2016, 0.9\% in 2017, and 2.0\% in 2018. Furthermore, the luminosity measurements in 2017-2018 have additional uncertainties, of 0.6 and 0.2\%, respectively, that are considered correlated between the two years. The overall uncertainty in integrated luminosity for the 2016--2018 period is 1.6\%.

\textbf{Pileup}: the uncertainty associated with the number of pileup interactions per bunch crossing is estimated by varying the total inelastic pp cross section by 4.6\%~\cite{Sirunyan:2018nqx}, fully correlated across the years.

\textbf{ECAL timing shift}: during the 2016--2017 data-taking periods, a gradual shift in the timing of the ECAL L1 trigger inputs occurred in the forward endcap region, $\abs{\eta} > 2.4$~\cite{CMS:2020cmk}. This led to a specific inefficiency due to erroneous association of detector readout to the previous bunch crossing in a small fraction of the collision events. A correction to this effect was determined using an unbiased data sample and found to be relevant in events containing high-\pt jets with $2.4<\abs{\eta} <3.0$. This correction is applied to simulation and is accompanied by a 20\% uncertainty. The uncertainty predominantly affects the VBF category in the \mmbb analysis, with a negligible effect on the results in this channel.

\textbf{Jet energy corrections}: the jet energy scale (JES) uncertainties include several sources parameterized as a function of the jet \pt and $\eta$~\cite{CMS:2016lmd}. Those variations can modify the content of the selected event sample. They also introduce event migration between categories. In the \mmbb analysis, the event \ptmiss changes as a result of variations in the jet kinematics whereas in the \ttbb analysis, JES uncertainties affect the \mtt distribution. Variations in the expected signal yield are between 15--50\% in the \mmbb analysis. In the \ttbb channel, distributions vary between 10--15\% of the nominal. Depending on the source, JES uncertainties are considered as uncorrelated, fully correlated, or partially correlated (50\%) across the years. The jet energy resolution is also considered, where the smearing corrections are varied within their uncertainties, uncorrelated across the years.

\textbf{\PQb tagging}: sources of systematic uncertainty that affect the data-to-simulation corrections of the \PQb tagging discriminant distribution are JES, the light flavor or gluon  (LF) jet contamination in the \PQb jet sample, the heavy flavor (HF) jet contamination in the LF jet sample, and the statistical fluctuations in data and MC~\cite{Sirunyan:2017ezt}. The JES variations in \PQb tagging are obtained together with the JES uncertainties on jet kinematics and follow the same correlation pattern across the years. The statistical components of the \PQb tagging uncertainties are uncorrelated while the rest are assumed correlated between different periods.

\textbf{Muon reconstruction}: the data-to-simulation correction factors for the muon tracking, reconstruction and selection efficiencies are estimated using a ``tag-and-probe'' method~\cite{tagandprobe} in DY data and simulated samples. These uncertainties include the pileup dependence of the correction factors and are correlated across the years since common procedural uncertainties are the dominant source. The requirements between the two analyses are slightly different, mainly because of a different impact parameter in $\PGt\to\PGm$ decays. The corrections, and therefore associated systematic uncertainties, are applied in bins of muon \pt and $\abs{\eta}$ in the \mmbb analysis~\cite{Sirunyan:2018fpa}. In the \ttbb analysis, a 2\% uncertainty, independent of \pt and $\eta$, per muon is used~\cite{CMS:2021gxc} and treated as uncorrelated between simulated and $\tau$-embedded events. The muon momentum scale varies within 0.4--2.7\%~\cite{Sirunyan:2018fpa} and is accounted for in systematic uncertainties on the signal and background \mtt distribution. Its impact is found to be negligible in the \mmbb analysis.

\textbf{Electron reconstruction}: the electron energy scale uncertainties in \emu and \etau final states are accounted for using methods outlined in Ref.~\cite{Khachatryan:2015hwa}. The reconstruction and selection efficiencies are accompanied by a 2\% uncertainty per electron, independent of \pt and $\eta$~\cite{Electrons:2021}. Similar to muons, these uncertainties are uncorrelated between simulated and $\tau$-embedded events. Uncertainties in the electron energy scale also affect the shapes of the \mtt distributions and are accounted for.

\textbf{Hadronically decaying \PGt lepton reconstruction}: in \mutau and \etau\ final states, there are uncertainties associated with \tauh identification efficiencies and energy scale corrections where the variations depend on \pt(\tauh) and decay mode, ranging from 3--5\% and 0.2--1.1\%, respectively.
Systematic variations in the selected event yields as well as in the shapes of the distributions are taken into account.
Uncertainties are considered uncorrelated across the bins of \pt(\tauh) and different years for the MC~\cite{CMS:2018jrd}.
Uncertainties of the same source are treated as 50\% correlated between the embedded DY background and simulated samples.
For events with a genuine \tauh lepton matched at the generator level, energy scale uncertainties are considered using shape variations.
In the case of muons and electrons misidentified as \tauh candidates, energy scale corrections are applied in bins of \pt, $\eta$, and decay mode of the misidentified \tauh.
These corrections are associated with uncertainties.
A 50\% correlation is considered between the embedded and MC samples for these lepton energy scale uncertainties.

\textbf{Trigger efficiencies}: an uncertainty of 1\% is assigned to the HLT efficiency in the \mmbb analysis. In the \ttbb channel, an uncertainty of 2\% is applied per single-lepton trigger and 5--10\% on the dilepton triggers with a \tauh requirement. Uncertainties associated with trigger efficiencies affect the shape of the distributions in this channel. The shape effects are taken into account in both simulated and embedded backgrounds, where a 50\% correlation is considered between the two.

\textbf{Background estimations in} $\mathbf{\ttbb}$ \textbf{final state}: the \PZ boson \pt reweighting uncertainty in DY samples, which amounts to 10\% of the nominal value, is taken as a \mtt shape uncertainty. The embedded samples include a 4\% normalization uncertainty~\cite{embedding}.
Moreover, shape uncertainties related to tracking efficiencies and contamination from non-DY events in the embedded sample are considered. Since the contribution of the QCD multijet background in the \emu\ channel is obtained from a same-sign sideband region with a limited number of events, the validity of the method is tested in independent same-sign SBs. This test results in a 20\% normalization uncertainty.
The uncertainty in the scale factor between the same-sign SBs and opposite-sign SRs is modeled using shape variations in the fit used to obtain the nominal values.
The misidentification probability, $f$, of a jet as a \tauh candidate depends on the jet multiplicity.
A 20\% normalization uncertainty is applied to the estimate of the \wjets\ and QCD multijet backgrounds due to $f$ being measured in $\PZ\to\PGm\PGm$ events with different jet multiplicities.
In addition, shape variations due to different measurements of $f$ are considered.

\textbf{Limited size of the samples}: to account for the limited size of the simulated samples, as well as the data in SBs used to estimate backgrounds, a bin-by-bin statistical uncertainty is considered where a Poisson nuisance parameter per bin is assigned to distributions in those samples~\cite{Barlow:249779}. This uncertainty is specific to the \ttbb analysis.

\textbf{Modeling uncertainties}: a total uncertainty of 3.6\% is assigned to the sum of the ggF and VBF Higgs boson production cross sections~\cite{deFlorian:2016spz} predicted by the SM and used to describe the upper limits on  $\mathcal{B}(\Hb\to\ab\ab\to \mmbb/\ttbb)$. It includes uncertainties from the perturbative QCD calculations, PDFs, and \alpS. In the \ttbb analysis, PDF and \alpS uncertainties are considered for simulated backgrounds, namely: 4.2\% for {\toppair}+jets, 5\% for Diboson, and 5\% for single top quark processes. These uncertainties are obtained following the PDF4LHC prescription~\cite{Butterworth_2016}. To account for variations in the signal acceptance in both channels,  the renormalization and factorization scales are doubled and halved simultaneously in simulation. In addition, the eigenvectors of the NNPDF3.1 PDF set are varied within their uncertainties in the final fit. The value of $\alpha_{\rm S}$, computed at the energy scale of the \PZ boson mass, is also varied within its uncertainty in the PDF set. For the parton shower simulation, uncertainties are separately assessed for initial- and final-state radiation, by varying the respective scales up and down by factors of two. Using the same model assumptions and procedures, the aforementioned uncertainties are considered fully correlated across the data-taking years.
\section{Results}
No excess of events over the expected SM backgrounds is observed in either of the \mmbb and \ttbb channels.
Upper limits are placed, at 95\% \CL, on $\mathcal{B}(\Hb\to\ab\ab\to \ell\ell \PQb\PQb)$ as a function of \ma, with $\ell$ being either a \PGt lepton or muon.
The two final states are combined to set upper limits on $\mathcal{B}(\Hb\to\ab\ab)$, assuming fixed decay fractions of \ab. The branching fraction $\mathcal{B}(\ab\to\mathrm{ff})$ depends on the 2HDM+S parameters, where $\mathrm{f}$ indicates either muon, \PQb quark, or \PGt lepton. Since the results in both channels are statistically limited, the combination mostly benefits from the additional data. The combined results are still dominated by the statistical uncertainties. At $\ma=35\GeV$, all systematic uncertainties amount to about 6\% of the total uncertainty, with the dominant contributions corresponding to JES in the \mmbb channel, followed by the theoretical uncertainties in the signal, and finally the uncertainties in the QCD multijet backgrounds in the \emu final state of the \ttbb analysis.

\begin{figure}[ht!]
  \centering
  \includegraphics[width=0.42\textwidth]{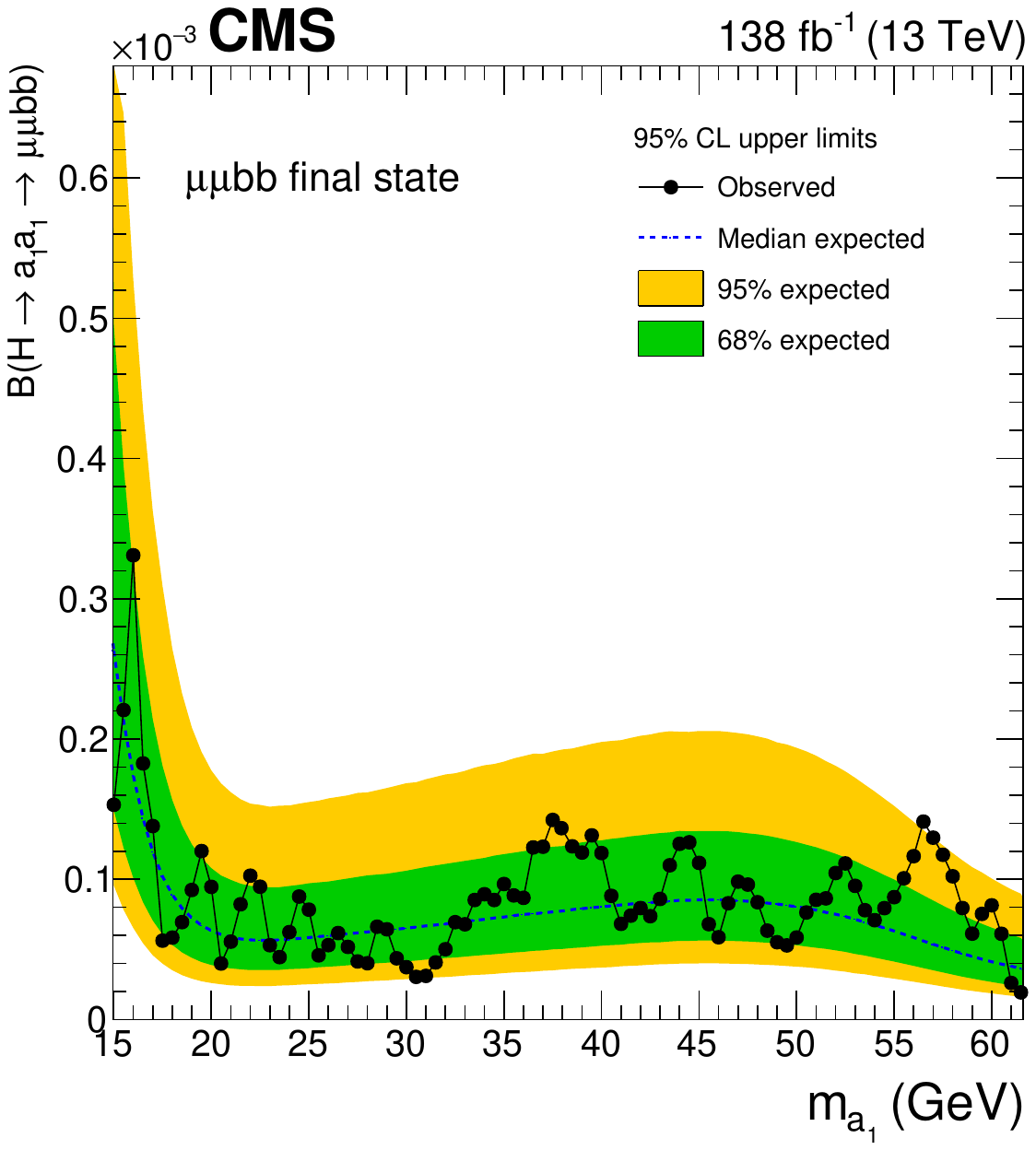}
  \caption{Observed and expected upper limits at 95\% \CL on $\mathcal{B}(\Hb\to\ab\ab\to \mmbb)$ as functions of \ma. The inner and outer bands indicate the regions containing the distribution of limits located within 68 and 95\% confidence intervals, respectively, of the expectation under the background-only hypothesis.}
    \label{fig:finalmmbb}
\end{figure}

Figure~\ref{fig:finalmmbb} shows the upper limits on $\mathcal{B}(\Hb\to\ab\ab\to \mmbb)$ at 95\% \CL, assuming SM predictions for the Higgs boson production cross section.
The \mmbb search is optimized for \ma values between 15 and 60\GeV, with signal sensitivity falling rapidly below $\ma=20\GeV$. This is mainly because the two \PQb jets start to merge as a result of a higher momentum for \ab. At 95\% \CL, the observed upper limits are (0.17--3.3)$\times 10^{-4}$ for the mass range 15 to 62.5\GeV, while the expected limits are (0.35--2.6)$\times 10^{-4}$.

\begin{figure*}[ht!]
  \centering
  \includegraphics[width=0.42\textwidth]{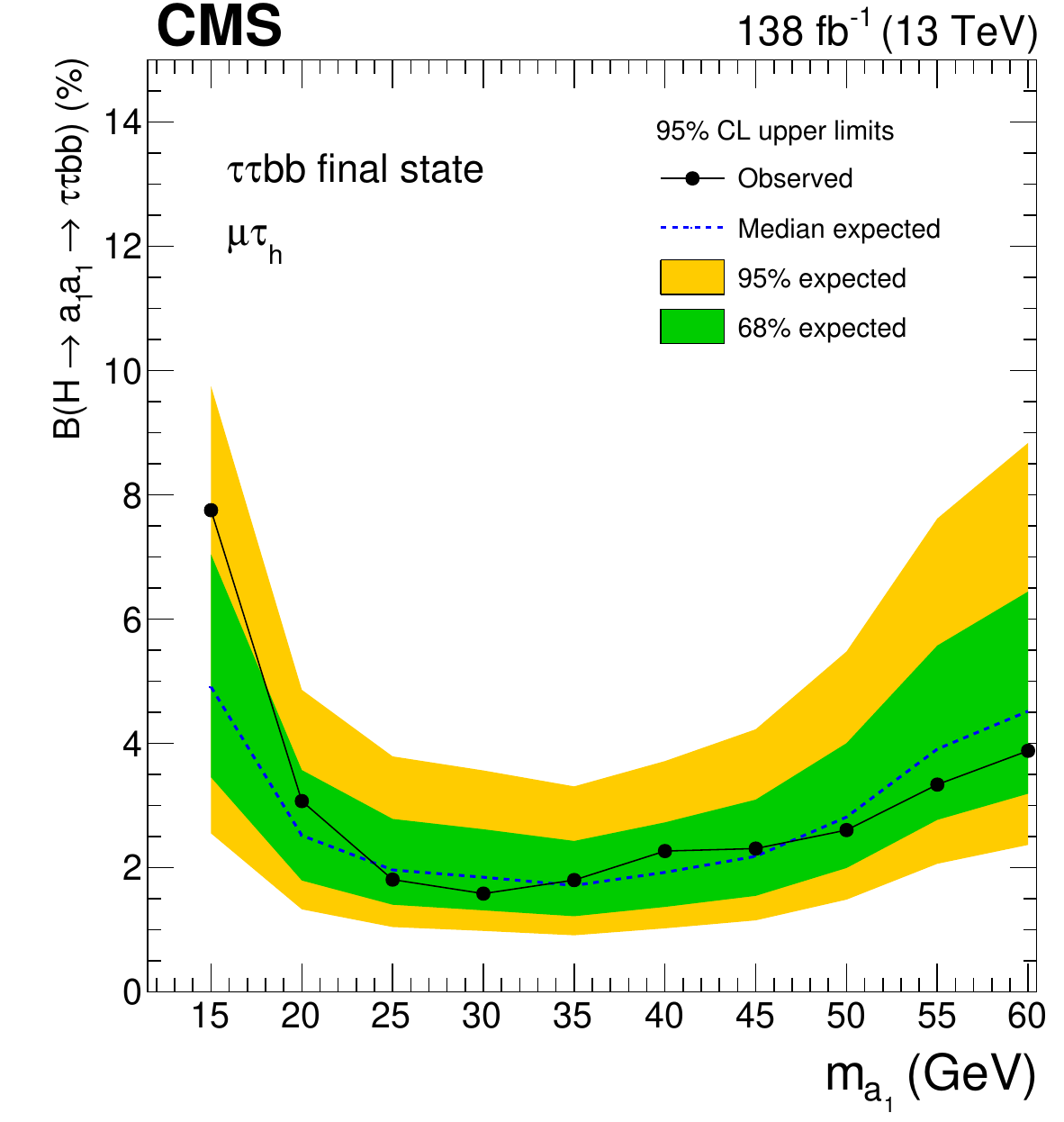}
  \includegraphics[width=0.42\textwidth]{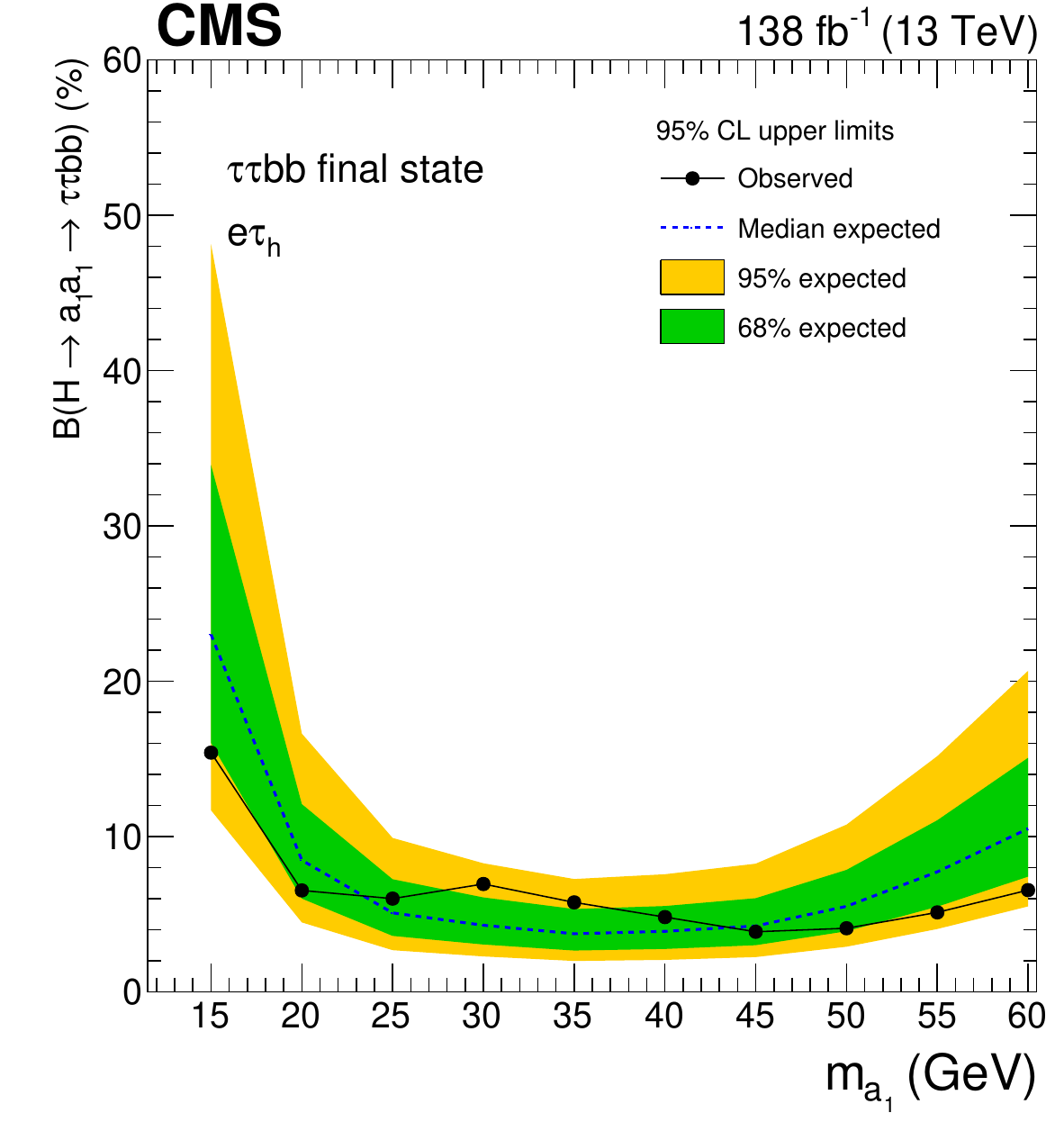} \\
  \includegraphics[width=0.42\textwidth]{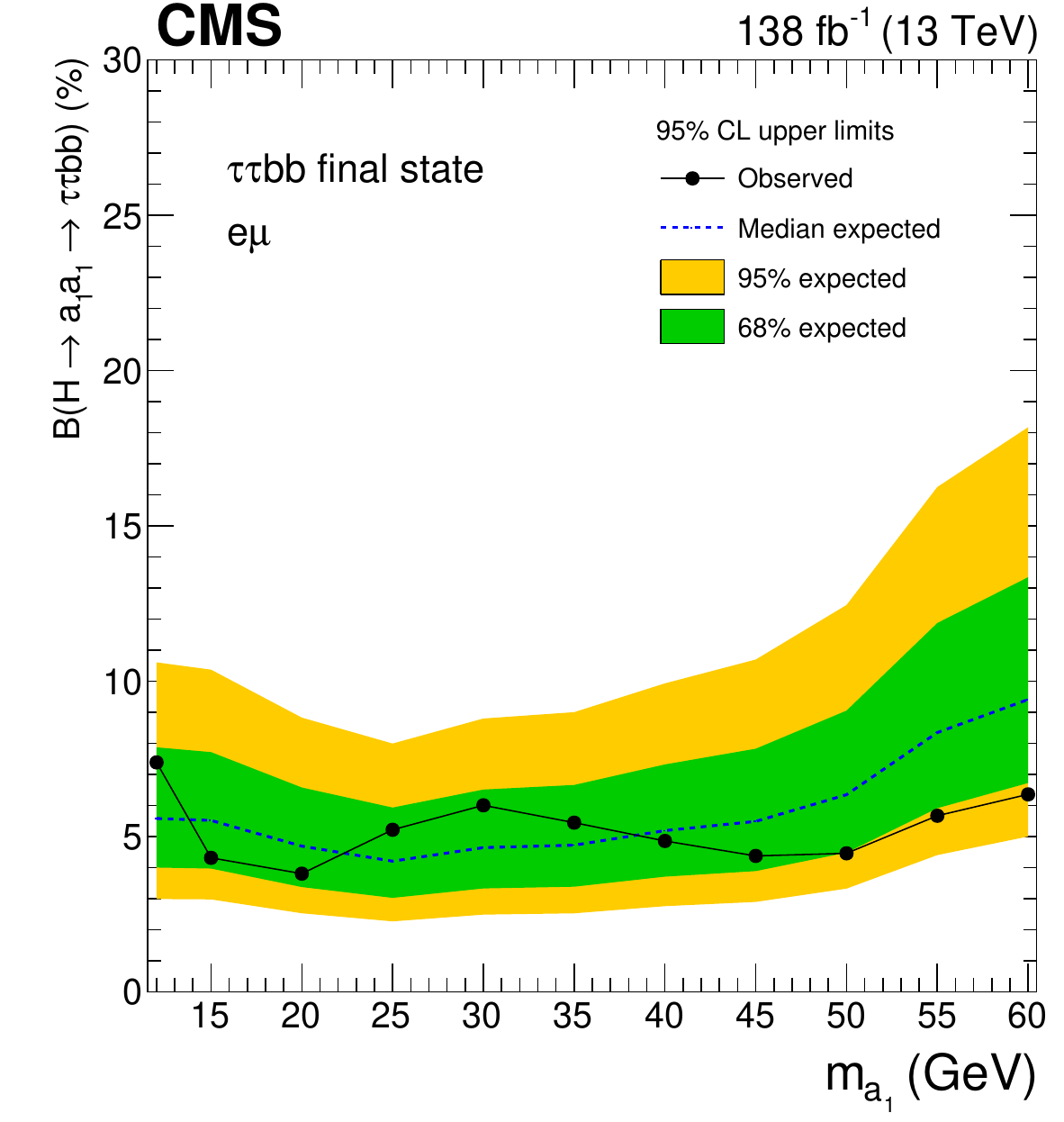}
  \includegraphics[width=0.42\textwidth]{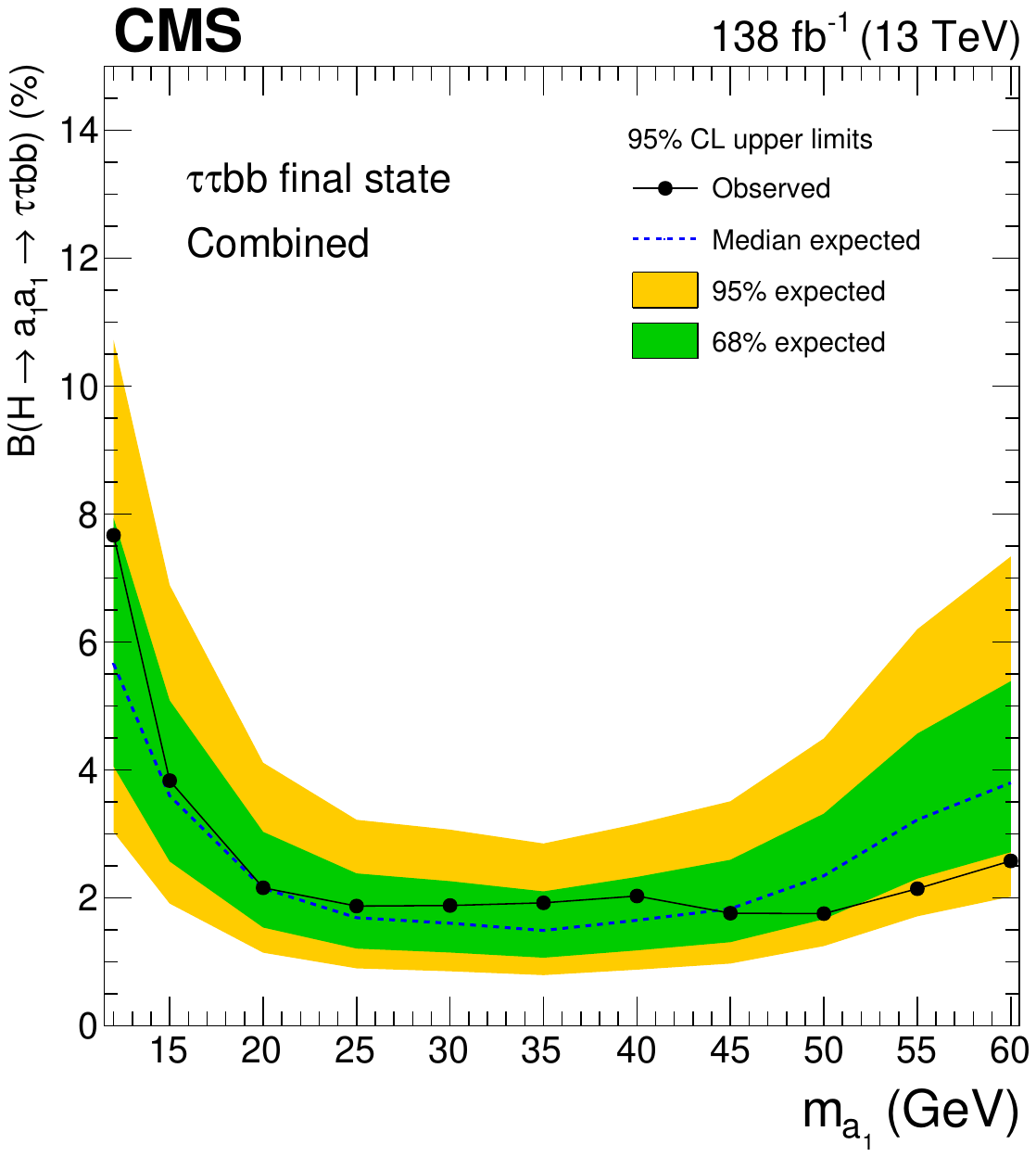}
  \caption{Observed and expected 95\% \CL exclusion limits on $\mathcal{B}(\Hb\to\ab\ab\to\ttbb)$ in percent, for the (upper left) \mutau, (upper right) \etau, (lower left) \emu channels, and (lower right) for the combination of all the channels.}
    \label{fig:results_limits_bbtautau}
\end{figure*}

Figure~\ref{fig:results_limits_bbtautau} shows the observed and expected 95\% \CL upper limits on $\mathcal{B}(\Hb\to\ab\ab\to\ttbb)$ as functions of \ma.
Only the \emu\ channel provides sensitivity to the 12\GeV mass point, as in this channel the baseline selection on the $\Delta R$ between the two \PGt candidates is the lowest.
For small \ma values, the decay products appear as boosted and may not be reconstructed as two separate objects.
The low $\Delta R$ requirement allows a selection of more signal events where the two \PGt candidates are close to each other.
The \mutau final state is the most sensitive, where limits as low as around 1.8\% (1.7\%) are observed (expected) in the intermediate mass range at $\ma= 35\GeV$.
Combining all final states in the \ttbb channel, observed limits on the branching fraction are found to be in the range 1.7--7.7\%, for a pseudoscalar mass between 12 and 60\GeV, with corresponding expected limits in the range 1.5--5.7\% at 95\% \CL.

\begin{figure}[ht!]
  \centering
  \includegraphics[width=0.42\textwidth]{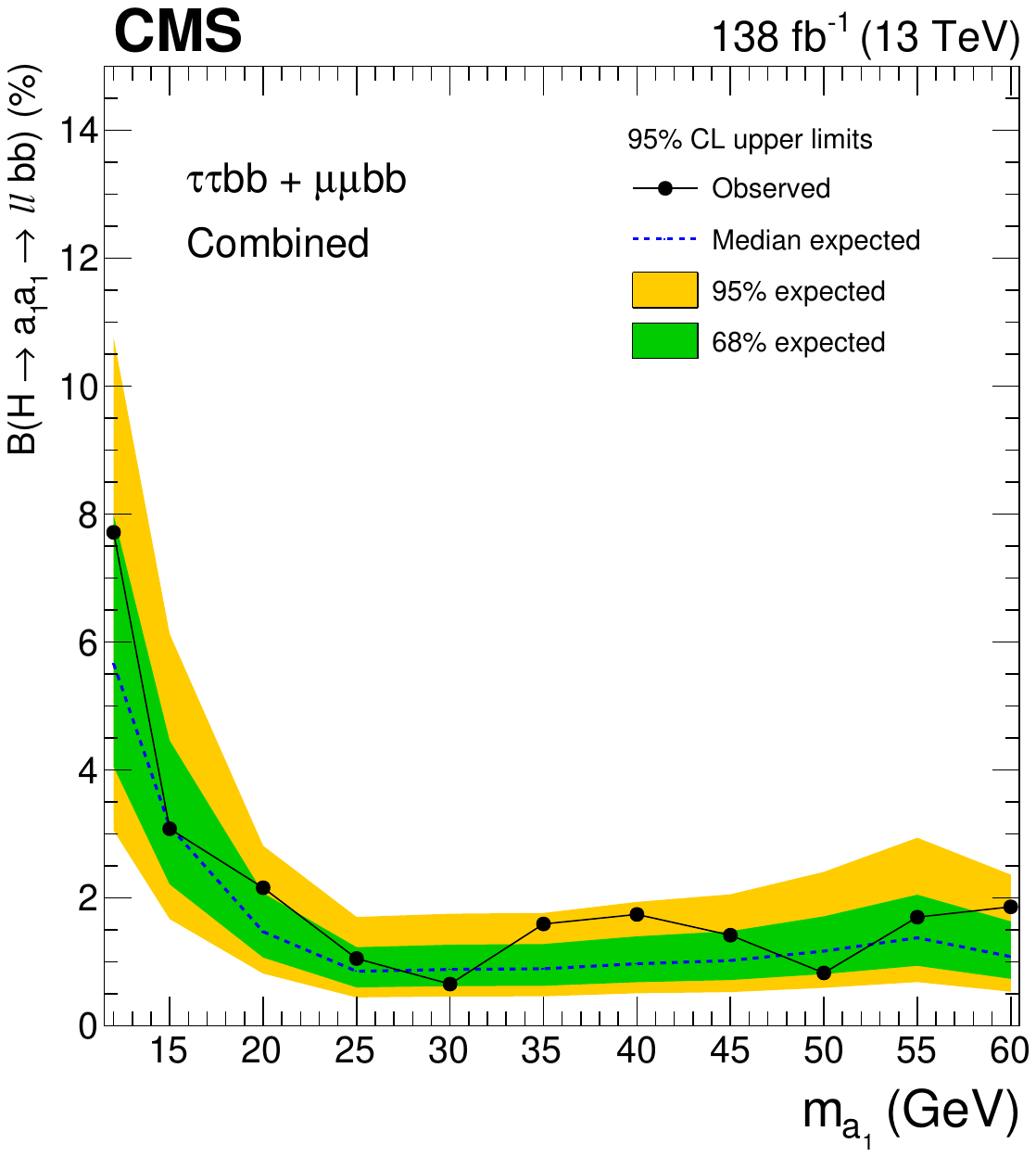}
  \caption{Observed and expected 95\% CL upper limits on $\mathcal{B}(\Hb\to\ab\ab\to \ell\ell\PQb\PQb)$ in \%, where $\ell$ stands for muons or \PGt leptons, obtained from the combination of the \mmbb and \ttbb channels. The results are obtained as functions of \ma for 2HDM+S models, independent of the type and $\tanb$ parameter. }
    \label{fig:results_limits_combined}
\end{figure}

Figure~\ref{fig:results_limits_combined} shows the observed and expected limits at 95\% \CL on $\mathcal{B}(\Hb\to\ab\ab\to \ell\ell\PQb\PQb)$, where $\ell$ stands for muons or \PGt leptons.
Using decay width expression from Ref.~\cite{Haisch:2018kqx}, the signal strength of each channel is scaled with a type and $\tanb$ independent factor to obtain this limit in the context of 2HDM+S models.
The observed and expected ranges are 0.6--7.7\% and 0.8--5.7\% respectively, depending on \ma.

\begin{figure}[ht!]
  \centering
  \includegraphics[width=0.42\textwidth]{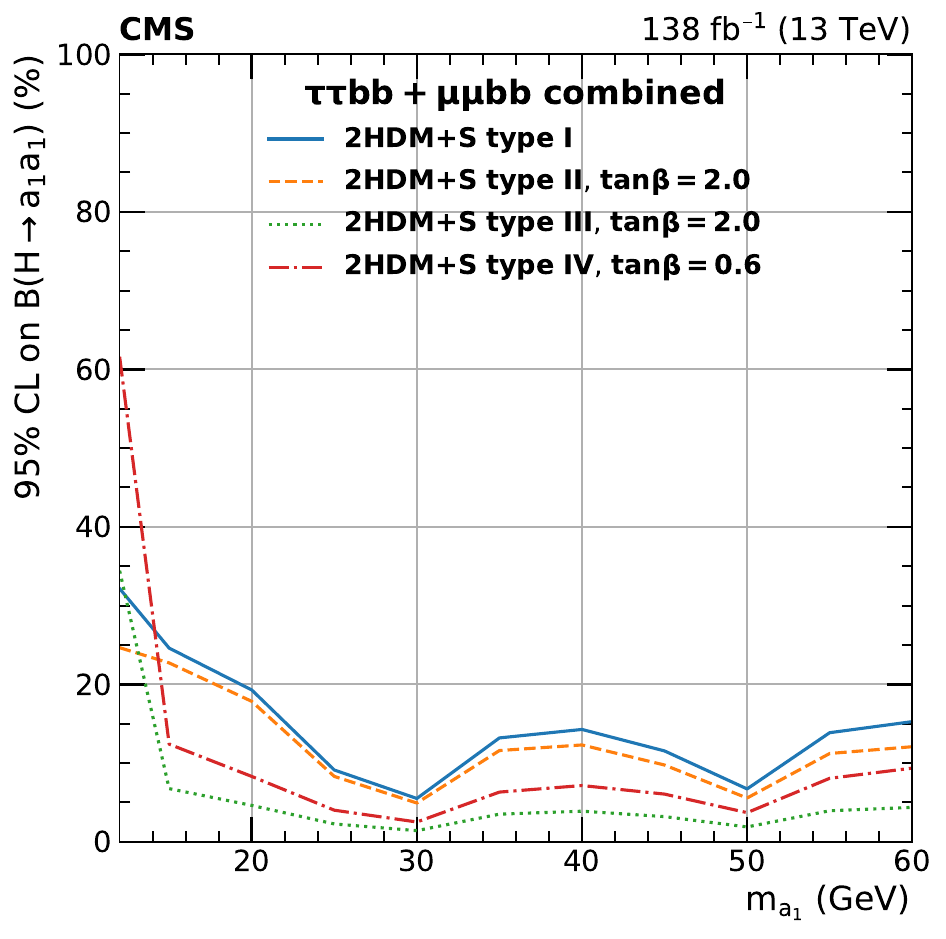}
  \caption{Observed and expected 95\% CL upper limits on $\mathcal{B}(\Hb\to\ab\ab)$ in \%, obtained from the combination of the \mmbb and \ttbb channels. The results are obtained as functions of \ma for 2HDM+S Type I (independent of $\tanb$), Type II ($\tanb=2.0$), Type III ($\tanb=2.0$), and Type IV ($\tanb=0.6$), respectively. }
    \label{fig:results_limits_type}
\end{figure}

The combined branching fraction $\mathcal{B}(\Hb\to\ab\ab)$ is obtained upon reinterpretation of the \mmbb and \ttbb results in different types of 2HDM+S and $\tanb$ values for $15<\ma<60\GeV$, illustrated in Fig.~\ref{fig:results_limits_type}. Upper limits in the range 5--23\% are observed at 95\% \CL for all Type II scenarios with $\tanb > 1.0$.
The tightest constraint is obtained for the Type III scenario with $\tanb = 2.0$. At 95\% \CL, the observed upper limits on the combined branching fraction are in the range 1--7\%, with a similar range for the expected upper limits.
For the Type IV scenario, the observed upper limits on $\mathcal{B}(\Hb\to\ab\ab)$ at 95\% \CL are between about 3 and 15\% for $\tanb = 0.5$, with corresponding expected limits between about 3 and 11\%.

\begin{figure}[ht!]
  \centering
  \includegraphics[width=0.42\textwidth]{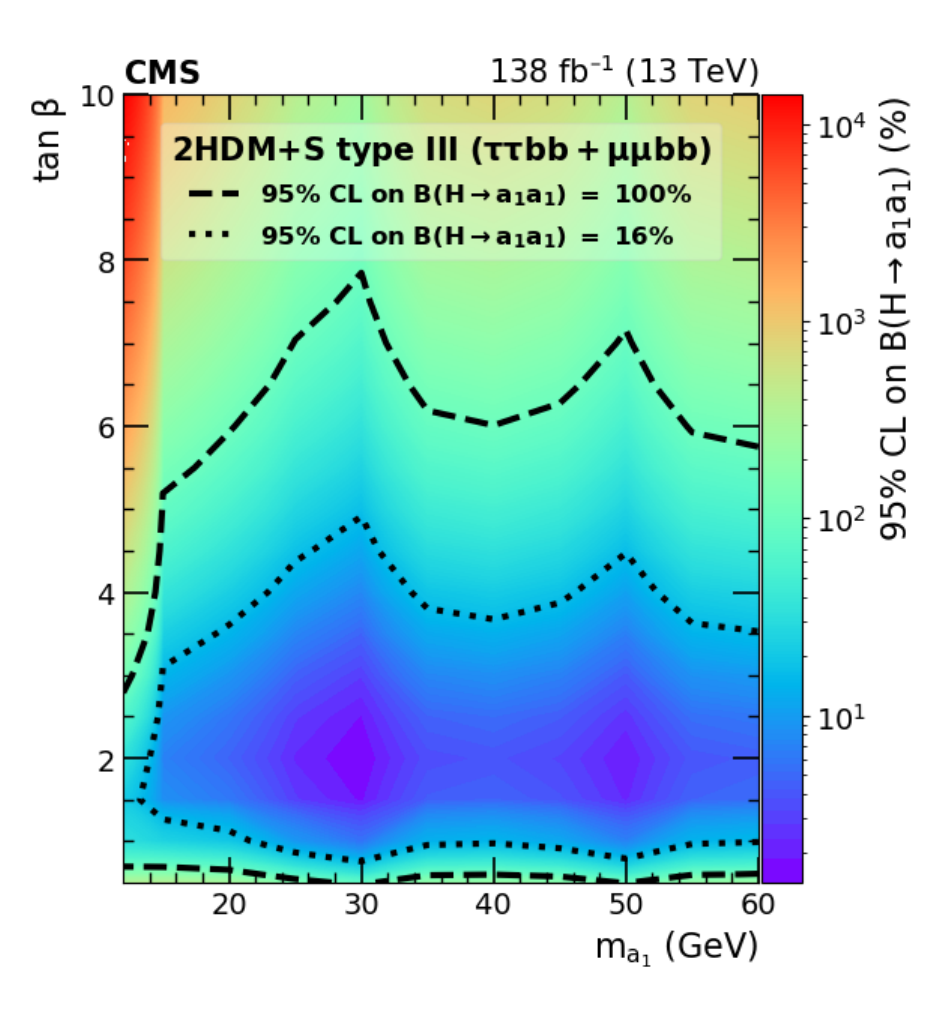}
  \includegraphics[width=0.42\textwidth]{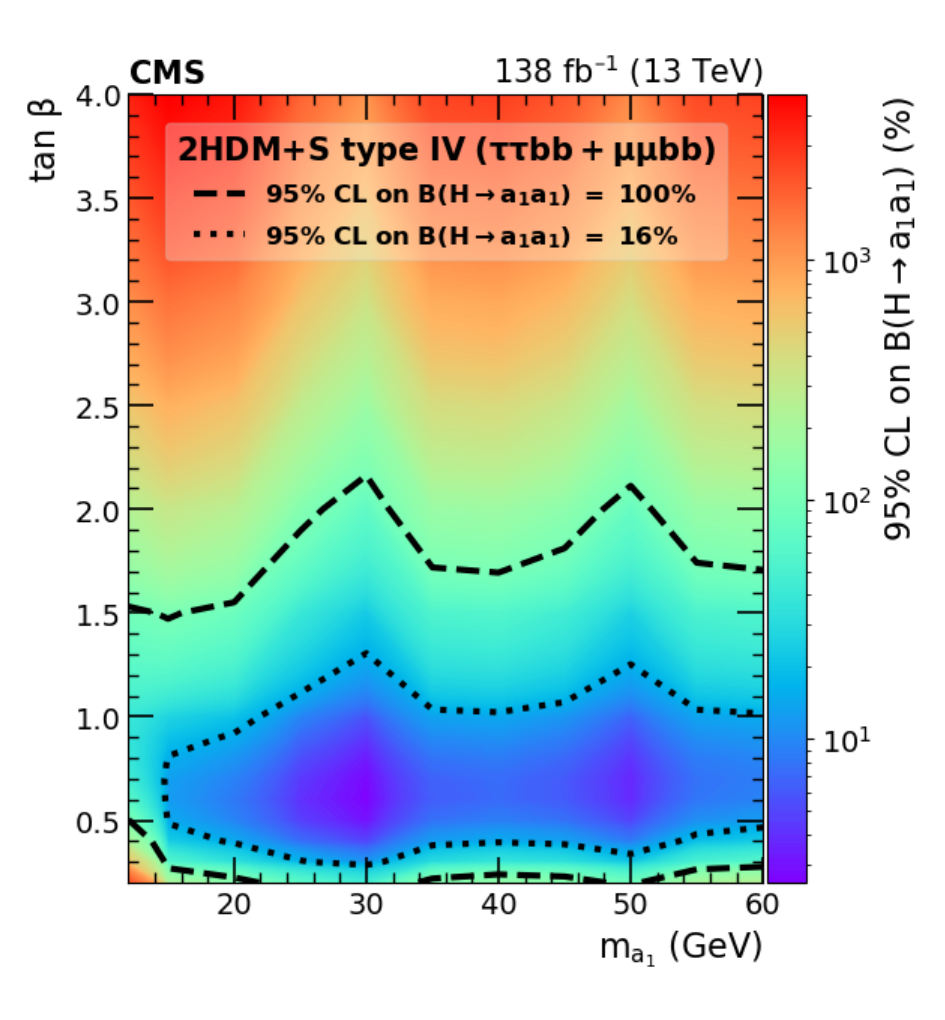}
  \caption{Observed 95\% CL upper limits on $\mathcal{B}(\Hb\to\ab\ab)$ in \%, for the combination of the \mmbb and \ttbb channels for Type III (\cmsLeft) and Type IV (\cmsRight) 2HDM+S in the \tanb \vs \ma parameter space.
The limits are calculated in a grid of 5\GeV in \ma and $0.1$--$0.5$ in \tanb, interpolating the points in between.
The contours corresponding to branching fractions of 100 and 16\% are drawn using dashed lines, where 16\% refers to the combined upper limit on Higgs boson to undetected particle decays from previous Run~2 results~\cite{naturecombcms}.
All points inside the contour are allowed within that upper limit.
}
    \label{fig:results_limits_combined_2D}
\end{figure}

The allowed values of \tanb and \ma are shown in Fig.~\ref{fig:results_limits_combined_2D} in the context of Type III and Type IV 2HDM+S.
The dashed contours represent the upper limits at 95\% \CL on Higgs boson to pseudoscalar decays, assuming the branching fraction to be either 100 or 16\%.
Here 16\% corresponds to the combined upper limit on Higgs boson to BSM particle decays obtained from previous Run~2 results~\cite{naturecombcms}.

\ifthenelse{\boolean{cms@external}}{}{\clearpage}
\section{Summary}
A search for an exotic decay of the 125\GeV Higgs boson (\PH) to a pair of light pseudoscalar bosons (\ab) in the final state with two b quarks and two muons or two $\PGt$ leptons has been presented. The results are based on a data sample of proton-proton collisions corresponding to an integrated luminosity of \il, accumulated by the CMS experiment at the LHC during Run~2 at a center-of-mass energy of 13\TeV. Final states with at least one leptonic \PGt decay are studied in the \ttbb channel, excluding those with two muons or two electrons. The results show significant improvement, with respect to the earlier CMS analyses at 13\TeV, beyond what is merely expected from the increase in the size of the data sample. A more thorough analysis of the signal properties using a single discriminating variable improves the \mmbb analysis, while the \ttbb analysis gains from a deep neural network based signal categorization. No significant excess in the data over the standard model backgrounds is observed. Upper limits are set, at 95\% confidence level, on branching fractions $\mathcal{B}(\Hb\to\ab\ab\to\mmbb)$ and $\mathcal{B}(\Hb\to\ab\ab\to\ttbb)$, in the \mmbb and \ttbb analyses, respectively. Both analyses provide the most stringent expected limits to date. In the \mmbb channel, the observed limits are in the range (0.17--3.3)$\times 10^{-4}$ for a pseudoscalar mass, \ma, between 15 and 62.5\GeV. Combining all final states in the \ttbb channel, limits are observed to be in the range 1.7--7.7\% for \ma between 12 and 60\GeV.
By combining the \mmbb and \ttbb channels, upper limits are set on the branching fraction $\mathcal{B}(\Hb\to\ab\ab\to \ell\ell\PQb\PQb)$, where $\ell$ stands for muons or \PGt leptons.  The observed upper limits range between 0.6 and 7.7\% depending on the \ma.  The results can also be interpreted in different types of 2HDM+S models. For \ma values between 15 and 60\GeV, $\mathcal{B}(\Hb\to\ab\ab)$ values above 23\% are excluded, at 95\% confidence level, in most of the Type II scenarios. In Types III and IV, observed upper limits as low as 1 and 3\% are obtained, respectively, for $\tanb=2.0$ and $0.5$. 

\begin{acknowledgments}
We congratulate our colleagues in the CERN accelerator departments for the excellent performance of the LHC and thank the technical and administrative staffs at CERN and at other CMS institutes for their contributions to the success of the CMS effort. In addition, we gratefully acknowledge the computing centers and personnel of the Worldwide LHC Computing Grid and other centers for delivering so effectively the computing infrastructure essential to our analyses. Finally, we acknowledge the enduring support for the construction and operation of the LHC, the CMS detector, and the supporting computing infrastructure provided by the following funding agencies: SC (Armenia), BMBWF and FWF (Austria); FNRS and FWO (Belgium); CNPq, CAPES, FAPERJ, FAPERGS, and FAPESP (Brazil); MES and BNSF (Bulgaria); CERN; CAS, MoST, and NSFC (China); MINCIENCIAS (Colombia); MSES and CSF (Croatia); RIF (Cyprus); SENESCYT (Ecuador); ERC PRG, RVTT3 and MoER TK202 (Estonia); Academy of Finland, MEC, and HIP (Finland); CEA and CNRS/IN2P3 (France); SRNSF (Georgia); BMBF, DFG, and HGF (Germany); GSRI (Greece); NKFIH (Hungary); DAE and DST (India); IPM (Iran); SFI (Ireland); INFN (Italy); MSIP and NRF (Republic of Korea); MES (Latvia); LMTLT (Lithuania); MOE and UM (Malaysia); BUAP, CINVESTAV, CONACYT, LNS, SEP, and UASLP-FAI (Mexico); MOS (Montenegro); MBIE (New Zealand); PAEC (Pakistan); MES and NSC (Poland); FCT (Portugal); MESTD (Serbia); MCIN/AEI and PCTI (Spain); MOSTR (Sri Lanka); Swiss Funding Agencies (Switzerland); MST (Taipei); MHESI and NSTDA (Thailand); TUBITAK and TENMAK (Turkey); NASU (Ukraine); STFC (United Kingdom); DOE and NSF (USA).

\hyphenation{Rachada-pisek} Individuals have received support from the Marie-Curie program and the European Research Council and Horizon 2020 Grant, contract Nos.\ 675440, 724704, 752730, 758316, 765710, 824093, 101115353, and COST Action CA16108 (European Union); the Leventis Foundation; the Alfred P.\ Sloan Foundation; the Alexander von Humboldt Foundation; the Science Committee, project no. 22rl-037 (Armenia); the Belgian Federal Science Policy Office; the Fonds pour la Formation \`a la Recherche dans l'Industrie et dans l'Agriculture (FRIA-Belgium); the Agentschap voor Innovatie door Wetenschap en Technologie (IWT-Belgium); the F.R.S.-FNRS and FWO (Belgium) under the ``Excellence of Science -- EOS" -- be.h project n.\ 30820817; the Beijing Municipal Science \& Technology Commission, No. Z191100007219010 and Fundamental Research Funds for the Central Universities (China); the Ministry of Education, Youth and Sports (MEYS) of the Czech Republic; the Shota Rustaveli National Science Foundation, grant FR-22-985 (Georgia); the Deutsche Forschungsgemeinschaft (DFG), under Germany's Excellence Strategy -- EXC 2121 ``Quantum Universe" -- 390833306, and under project number 400140256 - GRK2497; the Hellenic Foundation for Research and Innovation (HFRI), Project Number 2288 (Greece); the Hungarian Academy of Sciences, the New National Excellence Program - \'UNKP, the NKFIH research grants K 124845, K 124850, K 128713, K 128786, K 129058, K 131991, K 133046, K 138136, K 143460, K 143477, 2020-2.2.1-ED-2021-00181, and TKP2021-NKTA-64 (Hungary); the Council of Science and Industrial Research, India; ICSC -- National Research Center for High Performance Computing, Big Data and Quantum Computing, funded by the NextGenerationEU program (Italy); the Latvian Council of Science; the Ministry of Education and Science, project no. 2022/WK/14, and the National Science Center, contracts Opus 2021/41/B/ST2/01369 and 2021/43/B/ST2/01552 (Poland); the Funda\c{c}\~ao para a Ci\^encia e a Tecnologia, grant CEECIND/01334/2018 (Portugal); the National Priorities Research Program by Qatar National Research Fund; MCIN/AEI/10.13039/501100011033, ERDF ``a way of making Europe", and the Programa Estatal de Fomento de la Investigaci{\'o}n Cient{\'i}fica y T{\'e}cnica de Excelencia Mar\'{\i}a de Maeztu, grant MDM-2017-0765 and Programa Severo Ochoa del Principado de Asturias (Spain); the Chulalongkorn Academic into Its 2nd Century Project Advancement Project, and the National Science, Research and Innovation Fund via the Program Management Unit for Human Resources \& Institutional Development, Research and Innovation, grant B37G660013 (Thailand); the Kavli Foundation; the Nvidia Corporation; the SuperMicro Corporation; the Welch Foundation, contract C-1845; and the Weston Havens Foundation (USA).
\end{acknowledgments}

\bibliography{auto_generated} 
\cleardoublepage \appendix\section{The CMS Collaboration \label{app:collab}}\begin{sloppypar}\hyphenpenalty=5000\widowpenalty=500\clubpenalty=5000
\cmsinstitute{Yerevan Physics Institute, Yerevan, Armenia}
{\tolerance=6000
A.~Hayrapetyan, A.~Tumasyan\cmsAuthorMark{1}\cmsorcid{0009-0000-0684-6742}
\par}
\cmsinstitute{Institut f\"{u}r Hochenergiephysik, Vienna, Austria}
{\tolerance=6000
W.~Adam\cmsorcid{0000-0001-9099-4341}, J.W.~Andrejkovic, T.~Bergauer\cmsorcid{0000-0002-5786-0293}, S.~Chatterjee\cmsorcid{0000-0003-2660-0349}, K.~Damanakis\cmsorcid{0000-0001-5389-2872}, M.~Dragicevic\cmsorcid{0000-0003-1967-6783}, A.~Escalante~Del~Valle\cmsorcid{0000-0002-9702-6359}, P.S.~Hussain\cmsorcid{0000-0002-4825-5278}, M.~Jeitler\cmsAuthorMark{2}\cmsorcid{0000-0002-5141-9560}, N.~Krammer\cmsorcid{0000-0002-0548-0985}, D.~Liko\cmsorcid{0000-0002-3380-473X}, I.~Mikulec\cmsorcid{0000-0003-0385-2746}, J.~Schieck\cmsAuthorMark{2}\cmsorcid{0000-0002-1058-8093}, R.~Sch\"{o}fbeck\cmsorcid{0000-0002-2332-8784}, D.~Schwarz\cmsorcid{0000-0002-3821-7331}, M.~Sonawane\cmsorcid{0000-0003-0510-7010}, S.~Templ\cmsorcid{0000-0003-3137-5692}, W.~Waltenberger\cmsorcid{0000-0002-6215-7228}, C.-E.~Wulz\cmsAuthorMark{2}\cmsorcid{0000-0001-9226-5812}
% ...
\par}
\vskip\cmsinstskip
\dag:~Deceased\\
$^{1}$Also at Yerevan State University, Yerevan, Armenia\\
$^{2}$Also at TU Wien, Vienna, Austria\\
% ...
\end{sloppypar}
\end{document}